\documentclass[printer]{aa}
\usepackage{natbib}
\bibpunct{(}{)}{;}{a}{}{,} 
\usepackage[utf8]{inputenc}
\usepackage{graphicx}
\usepackage[varg]{txfonts}
\usepackage{xcolor}
\usepackage[colorlinks=true, linkcolor=blue, citecolor=blue, filecolor=blue, urlcolor=blue]{hyperref}
\usepackage{amsmath}
\usepackage{lscape}
\newcommand{\sci}[2]{\ensuremath{#1\times10^{#2}}}
\newcommand{\exl}{\object{EX~Lupi}}
\newcommand{\macc}{\ensuremath{\dot{M}_{\mathrm{acc}}}}
\newcommand{\lacc}{\ensuremath{L_{\mathrm{acc}}}}

\begin{document} 
   \title{Brightness and mass accretion rate evolution during the 2022 burst of EX~Lupi\thanks{Based on observations collected at the European Southern Observatory under ESO programmes 085.C-0764, 108.23N8, and 109.24F7.}}
   \titlerunning{The 2022 burst of EX~Lupi}

   \author{F.~Cruz-Sáenz de Miera\inst{1}\fnmsep\inst{2}\fnmsep\inst{3}
          \and
          Á.~Kóspál\inst{1}\fnmsep\inst{2}\fnmsep\inst{4}\fnmsep\inst{5}
          \and
          P.~Ábrahám\inst{1}\fnmsep\inst{2}\fnmsep\inst{4}
          \and
          R.~A.~B.~Claes\inst{6}
          \and
          C.~F.~Manara\inst{6}
          \and
          J.~Wendeborn\inst{7}
          \and
          E.~Fiorellino\inst{8}
          \and
          T.~Giannini\inst{9}
          \and
          B.~Nisini\inst{9}
          \and
          A.~Sicilia-Aguilar\inst{10}
          \and
          J.~Campbell-White\inst{6}
          \and
          J.~M.~Alcalá\inst{8}
          \and
          A.~Banzatti\inst{11}
          \and
          Zs.~M.~Szabó\inst{1}\fnmsep\inst{2}\fnmsep\inst{12}\fnmsep\inst{13}
          \and
          F.~Lykou\inst{1}\fnmsep\inst{2}
          \and
          S.~Antoniucci\inst{9}
          \and
          J.~Varga\inst{1}\fnmsep\inst{2}
          \and
          M.~Siwak\inst{1}\fnmsep\inst{2}
          \and
          S.~Park\inst{1}\fnmsep\inst{2}
          \and
          Zs.~Nagy\inst{1}\fnmsep\inst{2}
          \and
          M.~Kun\inst{1}\fnmsep\inst{2}
          }

   \institute{Konkoly Observatory, Research Centre for Astronomy and Earth Sciences,
              Eötvös Loránd Research Network (ELKH), Konkoly-Thege Miklós út 15--17,
              1121 Budapest, Hungary\\
              \email{fernando.cruz-saenz @ irap.omp.eu}
              \and
              CSFK, MTA Centre of Excellence,
              Konkoly Thege Miklós út 15--17, 1121, Budapest, Hungary
              \and
              Institut de Recherche en Astrophysique et Plan\'etologie, Universit\'e de Toulouse, UT3-PS, OMP, CNRS, 9 av.\ du Colonel Roche, 31028 Toulouse Cedex 4, France
              \and
              ELTE E\"otv\"os Lor\'and University, Institute of Physics, P\'azm\'any P\'eter s\'et\'any 1/A, 1117 Budapest, Hungary
              \and
              Max Planck Institute for Astronomy, Königstuhl 17, 69117 Heidelberg, Germany
              \and
              European Southern Observatory, Karl-Schwarzschild-Strasse 2, 85748, Garching bei München, Germany
              \and
              Department of Astronomy \& Institute for Astrophysical Research, Boston University, 725 Commonwealth Avenue, Boston, MA 02215, USA
              \and
              INAF-Osservatorio Astronomico di Capodimonte, via Moiariello 16, 80131 Napoli, Italy
              \and
              INAF-Osservatorio Astronomico di Roma, Via di Frascati 33, 00078 Monte Porzio Catone, Italy
              \and
              SUPA, School of Science and Engineering, University of Dundee, Nethergate, Dundee DD1 4HN, UK
              \and
              Department of Physics, Texas State University, 749 N Comanche Street, San Marcos, TX 78666, USA
              \and
              Max Planck Institute for Radioastronomy, Auf dem Hügel 69, 53121, Bonn, Germany
              \and
              SUPA, School of Physics and Astronomy, University of St Andrews, North Haugh, St Andrews, KY16 9SS, UK
             }

  \date{Received May 31, 2023; accepted August 4, 2023}

  \abstract
  {\exl{} is the prototype by which EXor-type outbursts were defined. It has experienced multiple accretion-related bursts and outbursts throughout the last decades, whose study have greatly extended our knowledge about the effects of these types of events. This star experienced a new burst in 2022.}
  {We aim to investigate whether this recent brightening was caused by temporarily increased accretion or by a brief decrease in the extinction, and to study the evolution of the \exl{} system throughout this event.}
  {We used multi-band photometry to create color-color and color-magnitude diagrams to exclude the possibility that the brightening could be explained by a decrease in extinction. We obtained spectra using the X-shooter instrument of the Very Large Telescope (VLT) to determine the \lacc{} and \macc{} during the peak of the burst and after its return to quiescence using two different methods: empirical relationships between line luminosity and \lacc{}, and a slab model of the whole spectrum. We examined the 130 year light curve of \exl{} to provide statistics on the number of outbursts experienced during this period of time.}
  {Our analysis of the data taken during the 2022 burst confirmed that a change in extinction is not responsible for the brightening. Our two approaches in calculating the \macc{} were in agreement, and resulted in values that are two orders of magnitude above what had previously been estimated for \exl{} using only a couple of individual emission lines, thus suggesting that \exl{} is a strong accretor even when in quiescence. We determined that in 2022 March the \macc{} increased by a factor of 7 with respect to the quiescent level.
  We also found hints that even though the \macc{} had returned to almost its pre-outburst levels, certain physical properties of the gas (i.e.\ temperature and density) had not returned to the quiescent values.}
  {We found that the mass accreted during this three month event was 0.8 lunar masses, which is approximately half of what is accreted during a year of quiescence. We calculated that if \exl{} remains as active as it has been for the past 130 years, during which it has experienced at least 3 outbursts and 10 bursts, then it will deplete the mass of its circumstellar material in less than 160\,000\,yr.}


   \maketitle

\section{Introduction}
\begin{figure*}
  \centering
  \includegraphics[width=\linewidth]{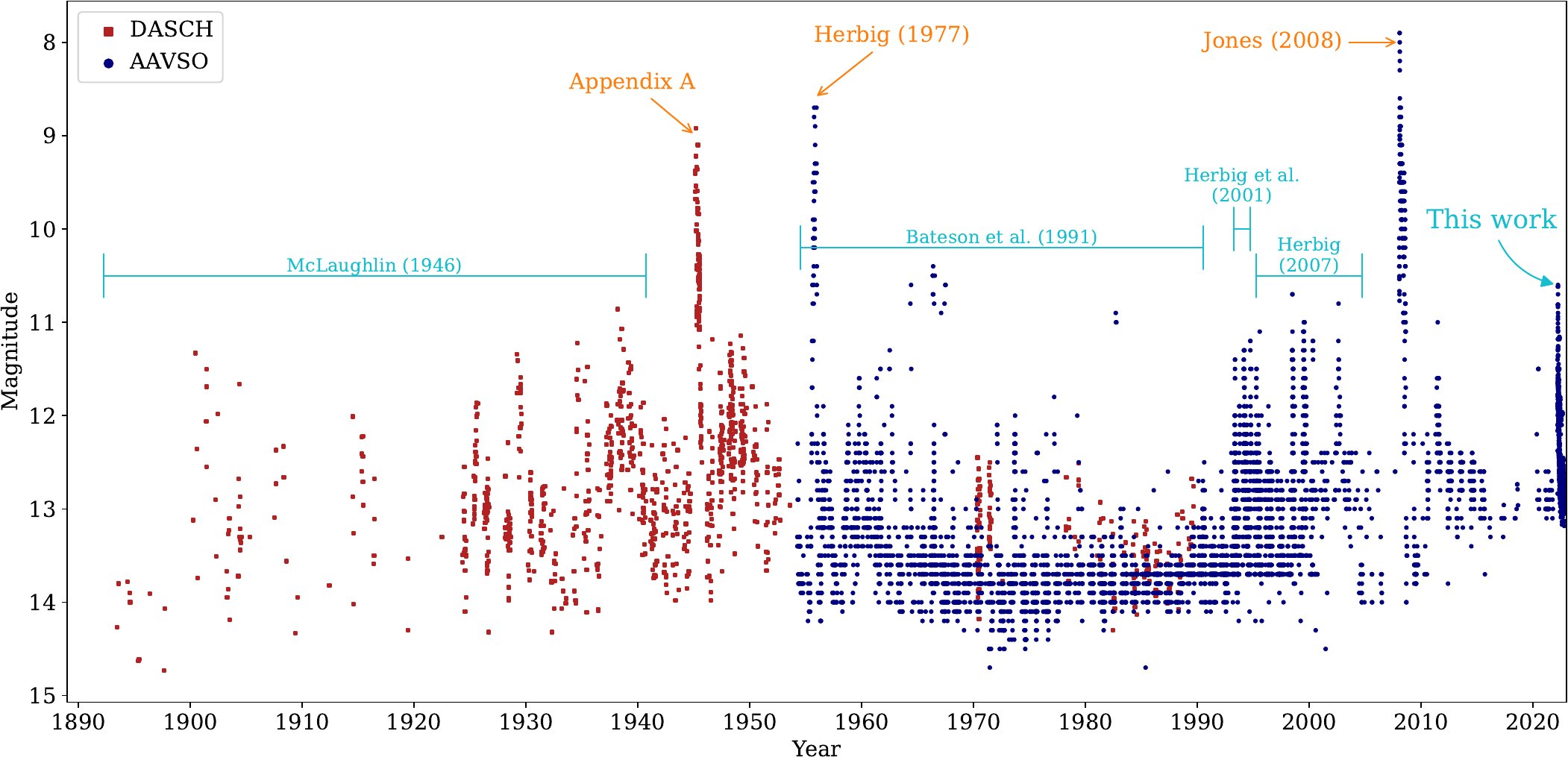}
  \caption{Light curve of \exl{} spanning the last $\sim$130 years. The photometry is a combination of Visible and $V$ magnitudes taken from DASCH (red markers) and AAVSO (blue markers). It shows three large outbursts in 1944 (see Appendix \ref{app:lc}), 1955 \citep{Herbig1977_ApJ217693H} and 2008 \citep{Jones2008_CBET12171J}, multiple bursts reported by \citet{McLaughlin1946_AJ52109M}, \citet{Bateson1991_PVSS1649B}, \citet{Herbig2001_PASP1131547H}, \citet{Herbig2007_AJ1332679H} and the one presented in this work (see Fig. \ref{fig:lightcurve}).\label{fig:full_lightcurve}}
\end{figure*}

Over the past 130 years, \exl{} has experienced at least three large outbursts (1944, 1955 and 2008) and more than a dozen smaller bursts (see Fig.\ \ref{fig:full_lightcurve} for the light curve and references).
This star has been taken to be the prototype of eruptive young stars that experience outbursts of 2--5 magnitudes lasting from a few months up to a year \citep[EXors,][]{Herbig1989_ESOC33233H}.
Together with their longer and more powerful counterparts (FUors), they represent the most dramatic cases of variability in low-mass young stellar objects \citep{Hartmann1996_ARAA34207H,Audard2014_prplconf387A,Fischer2023_PPVII}.
These events are caused by the sudden increase of the mass accretion rate (\macc{}) from $10^{-10}$--$10^{-8}$\,M$_\odot$\,yr$^{-1}$ in quiescence to $10^{-6}$--$10^{-4}$\,M$_\odot$\,yr$^{-1}$ during outburst, resulting in an increase of their bolometric luminosity by up to 2 orders of magnitude.
These accretion-related events can be detected via a $2-5$~mag brightening in optical and/or near-infrared photometric bands.
Recently, \citet{Fischer2023_PPVII} proposed an additional classification for the accretion-related events.
They classify as \textit{bursts} the events when a young star brightens by 1--2.5 magnitudes for at least a week and for as long as a year, and as \textit{outbursts} those which brighten by 2.5--6 magnitudes for many years to decades.

EXors can present both bursts and outbursts, and, due to their short durations, these events can occur repeatedly \citep{Herbig1989_ESOC33233H,Audard2014_prplconf387A}.
For example, \object{V1118~Ori} has experienced at least 5 outbursts in the last 30 years \citep{Giannini2020_AA637A83G}, \object{ASASSN-13db} has gone through one burst and one outburst since its discovery in 2013 \citep{SiciliaAguilar2017_AA607A127S}, \object{VY~Tau} experienced 14 outbursts between 1940 and 1970 \citep{Herbig1977_ApJ217693H}, and, as mentioned above, \exl{} has experienced multiple bursts and outbursts in the last 130 years (Fig.~\ref{fig:full_lightcurve}).
Indeed, the two latest identified EXors, namely Gaia19fct \citep{Hillenbrand2019_ATel133211H,Park2022_ApJ941165P} and Gaia20eae \citep{Ghosh2022_ApJ92668G,CruzSaenzdeMiera2022_ApJ927125C}, also show recurring bursts.

Historically, the discovery and monitoring of similar bursts in T~Tauri stars was rare and accidental.
However, in recent years the availability of all-sky monitoring or time-series surveys, operating at optical or infrared wavelengths have been very sensitive to such events.
The increasing statistics makes now possible further evaluation of the astrophysical significance of such bursts in the formation of low-mass stellar systems, such as their contribution to the build up of the protostar, and their impact on disk structure, material and chemistry.
Therefore, to properly evaluate their effects, we need to take into account that, while such bursts are less energetic than the outbursts, they have a shorter duty cycle and may produce significant cumulative effects that could impact the observations \citep{Claes2022_AA664L7C} and our overall interpretation of how disks evolve \citep{Manara2022_arXiv220309930M}.

In 2022 March, \citet{Zhou2022_ATel152711Z} and \citet{Kospal2022_RNAAS652K} reported that \exl{} was brightening again, based on $g$ band magnitudes from the All-Sky Automated Survey for Supernovae (ASAS-SN) survey, and on dedicated optical and near-infrared monitoring.
By 2022 March 10, the source had become 1.6\,mag brighter than the previous baseline in 2021.
An analysis of multi-band photometry taken during the early stages of the outbursts suggested that the brightening was not due to an decrease in the extinction but instead by an increase of the \macc{}.
We performed a multi-facility campaign using REM, LCOGT, VLT/X-shooter, VLTI/MATISSE, and ALMA observations to characterize the multi-band photometric evolution of the 2021 event to study the elevated mass accretion, and learn about the post-burst chemical and mineralogical impact on the disk.
Among the main astrophysical questions, our aim was to verify whether the magnetospheric accretion model is suitable to describe such bursts and outbursts, the possible role of instabilities in triggering the brightenings, and testing if the bursts would fit into a general picture that connects routine variability with the FUor-type outbursts \citep[e.g.][]{Liu2022_ApJ936152L}.
Here we report on the photometric monitoring and the UV/optical/near-infrared spectroscopic data sets, while the detailed analysis of the VLTI and ALMA measurements is planned for subsequent papers.
Two additional papers related to the accretion history of \exl{}, including the 2022 burst, are being prepared by \citet{Wang2023_ApJsubmitted} and \citet{SiciliaAguilar2023_MNRASsubmitted}.
The 2022 event of \exl{} is potentially one of the best studied burst in history, and, thus, it may be used as a prototype for bursts, as the 2008 event became the prototype of EXor outbursts.

We introduce \exl{} and its accretion history in Sect.\ \ref{sec:exlupi}.
In Sect.\ \ref{sec:obs}, we describe our observations and our calibration procedures.
We present the results and analysis of our photometric and spectroscopic observations in Sects.\ \ref{sec:photo} and \ref{sec:spec}, respectively.
In Sect.\ \ref{sec:disc} we discuss our findings and put the 2022 burst in context of other bursts and outbursts.
Finally, we summarize our work and list our conclusions in Sect.\ \ref{sec:theend}.

\section{EX~Lupi}\label{sec:exlupi}
\exl{} is the prototype of the EXor-type outbursts and, as such, it has been studied extensively by the astronomical community over the last decades.
It is a young stellar object located at a distance of 154.7$_{-0.7}^{+0.5}$\,pc \citep[Gaia DR3;][]{Gaia2016_AA595A1G,Gaia2022_arXiv220800211G}.
Its photospheric lines indicate that it has a radial velocity (RV) period of 7.417\,days \citep{Kospal2014_AA561A61K}. 
\citet{Alcala2017_AA600A20A} used an X-shooter spectrum to analyze \exl{} and, among their results, they found an extinction $A_V=1.1$.
Its first reported powerful ($>$5 mag) outburst began in 1955 \citep{Herbig1977_ApJ217693H} and was not observed spectroscopically.
In 2008, \exl{} underwent its largest outburst ever observed, during which the disk-to-star accretion rate increased to at least \sci{2}{-7}\,M$_\odot$~yr$^{-1}$ \citep{Juhasz2012_ApJ744118J}, leading to a more than 5\,mag optical brightening.
This elevated \macc{} was confined to within the innermost 0.2--0.4\,au region \citep{Goto2011_ApJ7285G,Kospal2011_ApJ73672K}.
Indeed, the mass accretion proceeds through remarkably stable accretion columns during both quiescence and outburst \citep{SiciliaAguilar2012_AA544A93S,SiciliaAguilar2015_AA580A82S}, and there are multiple indications for complicated radial and azimuthal disk structure \citep[e.g.][]{Rigliaco2020_AA641A33R}.
Observations during this powerful outburst provided the first evidence for on-going crystallization of silicate grains on the disk surface \citep{Abraham2009_Natur459224A}, and follow-up observations showed evidence for vertical and radial mixing within the disk \citep{Juhasz2012_ApJ744118J,Abraham2019_ApJ887156A}.
Recent JWST/MIRI observations showed that these crystals are now close to the water snowline and, as such, are in a position to be incorporated into planetesimals \citep{Kospal2023_ApJ945L7K}.
Moreover, the infrared molecular spectra showed strong signatures of enhanced UV photo-chemistry in the inner disk during outburst, with increased OH emission (probably from photo-dissociation of H$_2$O) and disappearance of organic molecules previously observed in quiescence \citep{Banzatti2012_ApJ74590B}.
In addition to these extreme outbursts, \exl{} has undergone at least 19 bursts in the past 130 years \citep{McLaughlin1946_AJ52109M,Bateson1991_PVSS1649B,Lehmann1995_AA300L9L,Herbig2001_PASP1131547H,Herbig2007_AJ1332679H,Abraham2019_ApJ887156A}.

\section{Observations}\label{sec:obs}
\subsection{Optical and near-infrared photometry}\label{ss:obs_photometry}
\begin{figure*}
  \centering
  \includegraphics[width=\textwidth]{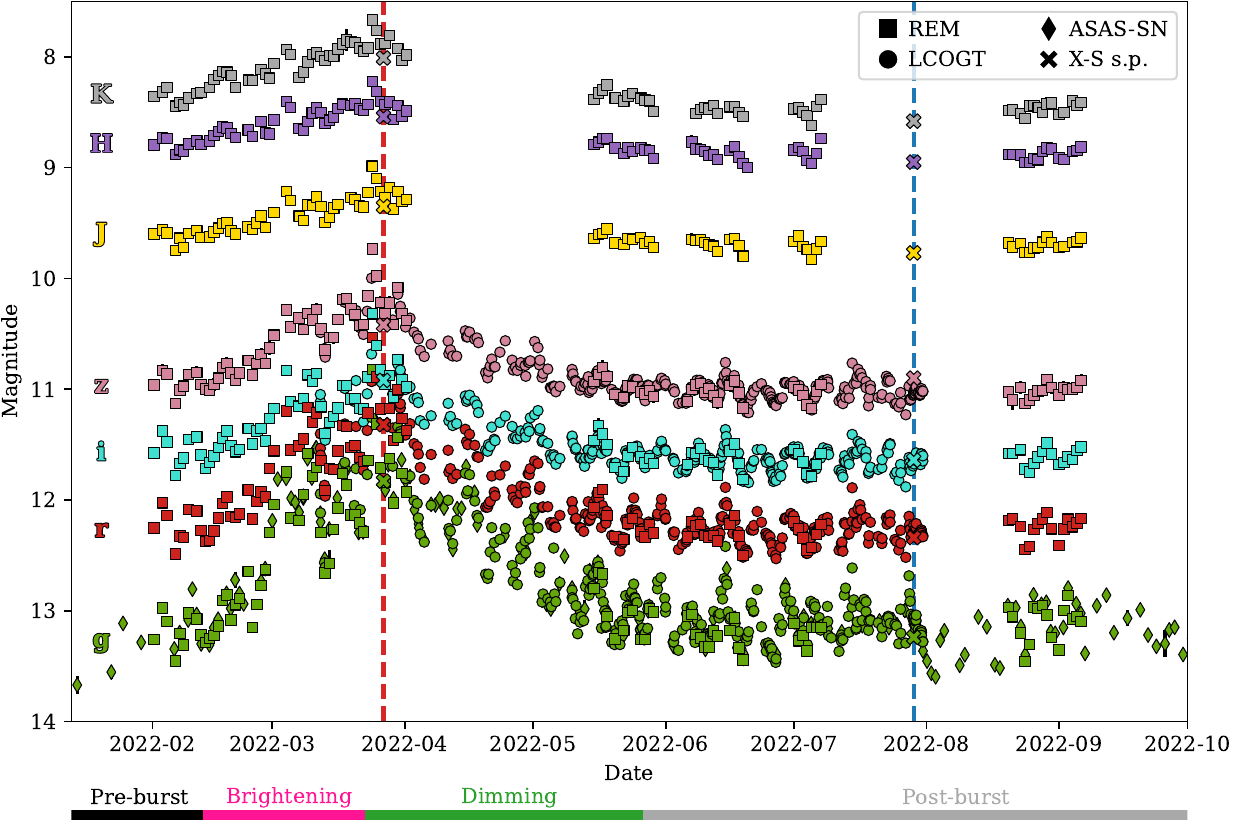}
  \caption{Light curve of \exl{} using optical and near-infrared (NIR) photometry, where each color represents a different filter as labeled on the left side of the plot. The square, circle and diamond symbols represent the different telescopes used to construct the light curve, and the X symbol indicates the synthetic photometry of the X-shooter spectra. The dates when these spectroscopic observations were taken are indicated in vertical dashed lines. The LCOGT and ASAS-SN magnitudes include the shifts mentioned in Sect.~\ref{ss:obs_photometry}. The colored line beneath the light curve indicates the start/end dates of the different stages of the light curve.\label{fig:lightcurve}}
\end{figure*}

We monitored \exl{} at optical and near-infrared wavelengths between 2022 February 1 and September 6 with the Rapid Eye Mount (REM) telescope\footnote{REM is a 60\,cm mirror diameter automatic telescope at La Silla (Chile) operated by the Italian National Institute for Astrophysics (INAF). \url{http://www.rem.inaf.it/}} (LT 44025; PI: E. Fiorellino).
We obtained images with an approximately nightly cadence using Sloan $g'r'i'z'$ and $JHK_s$ filters, except for several weeks in 2022 July -- August, when telescope operations had to be stopped due to an unusually strong and unprecedented snowstorm at La Silla.
We obtained aperture photometry for the target and ten comparison stars within 3$'$ of \exl{}.
We selected comparison stars with good quality APASS9 \citep{Henden2015_AAS22533616H} and 2MASS \citep{Cutri2003_yCat22460C} photometry and with sufficiently constant brightness ($\sigma$V $<$ 0.1\,mag).
We used an aperture radius of 3 pixel (or 3\farcs67) and a sky annulus between 9 and 12 pixels ($11''$ and $22''$).
Data from the first few weeks of our monitoring were published in \citet{Kospal2022_RNAAS652K}, while the full data set is shown in Fig.~\ref{fig:lightcurve}.
  
Additional $g'r'i'z'$ photometry of \exl{} was obtained between 2022 March 30 and 2022 July 31 using the Las Cumbres Observatory global network of telescopes (LCOGT).
Data sets were obtained with a roughly 5-hour cadence but were sporadic due to weather and schedule constraints.
LCOGT utilizes the BANZAI pipeline\footnote{\url{https://github.com/LCOGT/banzai}} \citep{BANZAI} to perform standard image reduction, astrometric correction, and photometry.
However, we found these to be inconsistent, particularly for images with low signal-to-noise ratio (SNR), thus, we performed astrometric correction on each frame using the Astrometry.net website\footnote{\url{https://astrometry.net/}}.
Frames for which no solution could be found were discarded, except when an image was obtained within a set so that the WCS information from a nearby corrected image could be used.
We selected all sources in the frame cross-matched with the ATLAS-REFCAT 2 survey\footnote{\url{https://archive.stsci.edu/hlsp/atlas-refcat2}} \citep{Tonry2018_ApJ867105T}, down to {$g'$}$\sim$16\,mag.
We pruned each frame from weak sources by carrying out aperture photometry using a 20-pixel wide aperture and a background annulus between 30 and 40 pixels, and discarding sources with with SNR$<$5.
Afterwards, we used least-squares to fit each source with a profile composed by a 2D Gaussian profile and a Moffat profile.
The flux of each source was computed as the total integral of the best-fit profile, while we kept the uncertainty obtained with the aforementioned annulus.
We then calculated a magnitude zero-point for each source using the flux estimated from our images and the magnitude from the ATLAS-REFCAT 2 catalog.
Magnitude zero-points were converted to flux zero points, 3$\sigma$ outlying flux zero-points were discarded and the average flux zero-point was then converted back to an average magnitude zero-point.
This was then used to calculate a calibrated apparent magnitude for each source in the frame, including \exl{}.
  
We complemented our photometric light curve with the $g$-band measurements obtained as part of the All-Sky Automated Survey for Supernovae (ASAS-SN) project\footnote{\url{https://www.astronomy.ohio-state.edu/asassn/index.shtml}} which monitors the full sky every night down to $g\sim$18\,mag \citep{Shappee2014_ApJ78848S,Kochanek2017_PASP129j4502K}.
The earliest date with ASAS-SN data for \exl{} in 2022 is January 14.
  
When we compared the magnitudes obtained with different facilities, we found that there are systematic differences. We corrected these by taking the median difference between the magnitudes of our REM program and the LCOGT and ASAS-SN data when we had observations on the same nights. In the case of the LCOGT data the differences were $+$0.144\,mag, $+$0.072\,mag, $+$0.006\,mag, and $-$0.200\,mag for the $g'r'i'z'$ filters, respectively, and for the ASAS-SN $g$ photometry we used a shift of +0.041\,mag.
We present the multi-filter light curve in Fig.~\ref{fig:lightcurve}, where these shifts have already been applied.

\subsection{X-shooter spectroscopy}\label{ss:xshooter}
\begin{figure*}
  \centering
  \includegraphics[width=\textwidth]{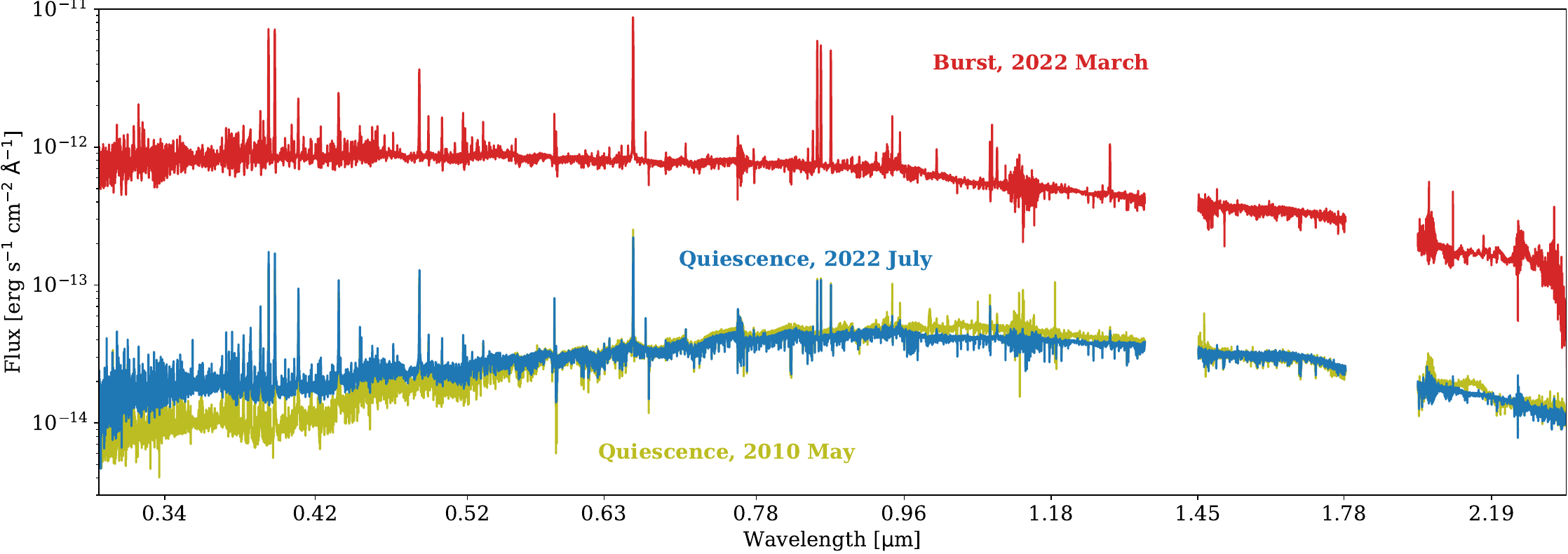}
  \caption{The three X-shooter spectra, including our two observations taken in 2022 and one archival epoch from 2010. It shows how the significant brightening of the whole spectrum due to the burst and how, after the burst ended, it had almost returned to its pre-burst shape.\label{fig:plot_all}}
\end{figure*}

We obtained spectra of \exl{} using the X-shooter échelle spectrograph \citep{Vernet2011_A&A536A105V} on ESO’s Very Large Telescope (VLT) on 2022 March 27 (108.23N8.001, PI: F.~Cruz-Sáenz de Miera) and 2022 July 29 (109.24F7.001, PI: F.~Cruz-Sáenz de Miera).
We complemented our spectroscopy data set with archival X-shooter observations taken on 2010 May 05 (085.C-0764, PI: H.M.~Günther) that were already published by \citet{Alcala2017_AA600A20A}.
We used the 0\farcs5, 0\farcs4, and 0\farcs4 wide slits, providing 9860, 18300, and 11400 spectral resolution in the UVB (0.30$-$0.56 $\mu$m), VIS (0.53$-$1.02 $\mu$m), and NIR (0.99$-$2.48 $\mu$m) arms, respectively.
In the case of the two 2022 observations, we also observed \exl{} with the 5$''$ slit on the three arms to correct for slit losses and, thus, have a reliable flux calibration.
The 2022 observations were carried out using an ABBA nodding pattern, while the 2010 ones were done with an AB pattern.
We reduced the raw data using the X-shooter pipeline (v.3.5.3) within the EsoReflex environment and we corrected for telluric absorption using ESO’s Molecfit \citep{Kausch2015_AA576A78K,Smette2015_AA576A77S}.
The results of our calibration for the NIR arm of the 2010 snapshot had an abnormal continuum shape, thus, for this arm, we used the calibrated product found in the ESO Science Archive\footnote{\url{http://archive.eso.org}} and then corrected it for telluric absorption as for our own data reduction.
As the final steps, we scaled each of the 2022 spectra so that their continuum levels matched those observed with the 5$''$ broad slit.
The 2010 observations did not include the broad slit as part of their setup, thus, we based our flux correction on a reprocessed version of the $VJHK$ REM photometry published by \citet{Kospal2014_AA561A61K} which, coincidentally, was taken just 2 hours before the X-shooter spectrum.
These observations were reprocessed for the present study, which led to higher flux densities for the X-shooter spectrum by 20\% in the UV and optical, while no change in the infrared.
We carried out synthetic photometry on the X-shooter spectrum and used the ratio between this and the observed photometry to find the scaling constants to correct the spectra.
We utilized the $V$ band to scale the UVB and VIS arms, and the $JHK$ bands to scale the NIR arm.
In the case of the latter, we fitted a straight line with the three ratios (and their respective wavelengths) to find the correction for the whole arm.
We present the three X-shooter spectra in Fig.~\ref{fig:plot_all}.

\section{Results and Analysis: Photometry}\label{sec:photo}
\subsection{Light curve}\label{ss:lightcurve}
In Fig.~\ref{fig:lightcurve} we present the light curve for \exl{} composed of different photometric filters at optical and near-infrared wavelengths.
Using the $g$ band to monitor the brightness levels of \exl{}, we divide the light curve into four stages: the ``pre-burst'' for all photometric points before 2022 February 14, the ``brightening'' stage between the previous date and when \exl{} reached its burst peak on March 28, the ``dimming'' phase began immediately after reaching the peak and finished when \exl{} went back to its pre-burst brightness level on May 23, and finally the "post-burst" stage for all points after this latter date.
The four stages are indicated in Fig.~\ref{fig:lightcurve} beneath the light curves.
We used these time intervals to determine the brightening and dimming rates, and we found that \exl{} brightened and dimmed at 0.036\,mag\,d$^{-1}$ and 0.026\,mag\,d$^{-1}$, respectively.

\subsection{Period analysis}
\begin{figure}
  \centering
  \includegraphics[width=\columnwidth]{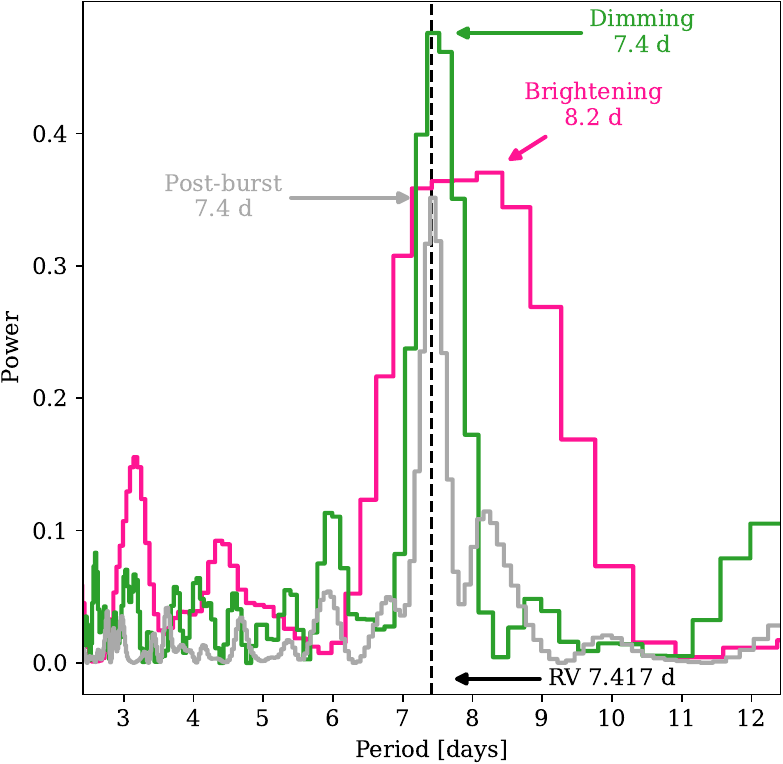}
  \caption{Lomb-Scargle periodograms for three of the light curve stages.
  They show that the periods during the dimming and post-burst stages are in agreement with the estimate by \citet{Kospal2014_AA561A61K} from radial velocity (RV) measurements.
  The periodogram of the brightening stage suggests that there was a change in the sinusoidal-like periodicity of \exl{}.\label{fig:periodogram}}
\end{figure}

A visual inspection of our photometric light curve indicates that \exl{} has periodic brightness fluctuations with a similar period to the rotational period of 7.417\,days found by \citet{Kospal2014_AA561A61K}.
To determine the periodicity of \exl{} in our data set, we produced Lomb-Scargle diagrams using only the $g$-band photometry and its uncertainties, as it has the best time coverage due to the combination of ASAS-SN, REM, and LCOGT photometry. The other bands exhibit similar periodicity characteristics, albeit with lower significance (owing to their poor coverage), and so are not considered in the rest of our periodicity analysis.
Our first step was to remove the burst from the light curve, which we did by fitting four straight lines covering the four stages previously mentioned (pre-burst, brightening, dimming, and post-burst), and subtracted the best-fit lines from each phase.
Afterwards, we produced a Lomb-Scargle diagram for each stage using the \texttt{LombScargle} function of Astropy \citep{astropy:2013, astropy:2018, astropy:2022}.
The pre-burst stage consists only of 20 data points, thus, its Lomb-Scargle diagram did not produce significant results and we did not analyze it further.
The periodograms for the remaining three stages are shown in Fig.~\ref{fig:periodogram}.

The dimming and post-outburst stages have periods of 7.4\,days and, thus, are in agreement with the estimate by \citet{Kospal2014_AA561A61K} of 7.417\,days within 0.5\,hours.
The periodogram of the brightening stage shows an almost flat peak, where the periods between 7.2 and 8.2\,days have quite similar probabilities.
An assumption of the Lomb-Scargle analysis is that the input light curve behaves sinusoidally, therefore, a non-sinusoidal behavior of the light curve during this phase could broaden the width of this distribution.

It is not clear what physical process could have changed the behavior of the light curve for only this phase.
However, we expect it to be related to changes in the accretion column such as the appearance of a second accretion column at a different position on the star, or the growth of the bright spot where the column makes contact with the star.
Understanding the nature behind these changes is beyond the scope of this paper, and, for now, we only conclude that the periodogram of the brightening phase should not be considered as a reliable deviation from the periodicity found by \citet{Kospal2014_AA561A61K}.

\subsection{Color-color and color-magnitude diagrams}
\begin{figure*}
  \centering
  \includegraphics[width=\textwidth]{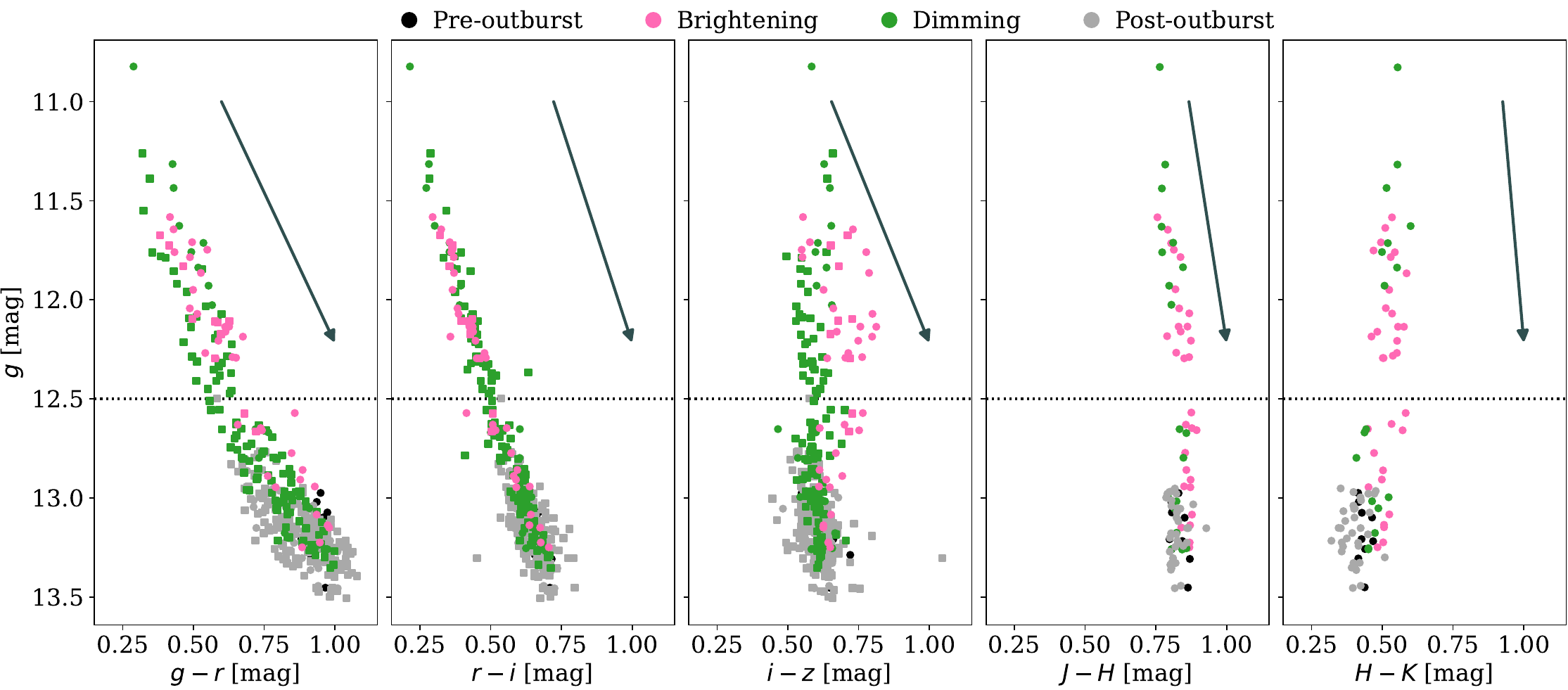}
  \caption{Color-magnitude diagrams for the optical and NIR photometry. The circle and square symbols are the REM and LCOGT observations, respectively. The colors indicate the different stages of the light curve where black is the pre-burst quiescence, pink is the brightening phase, green is the dimming phase and gray is the post-burst quiescence (cf.\ Sect.~\ref{ss:lightcurve}). The arrow is the $A_V$ = 1.1\,mag extinction vector following the extinction curve of \citet{Cardelli1989_ApJ345245C}. The dotted horizontal lines indicate the $g=12.5$\,mag threshold indicative of when \exl{} is in burst.\label{fig:cmd}}
\end{figure*}

\begin{figure}
  \centering
  \includegraphics[width=\columnwidth]{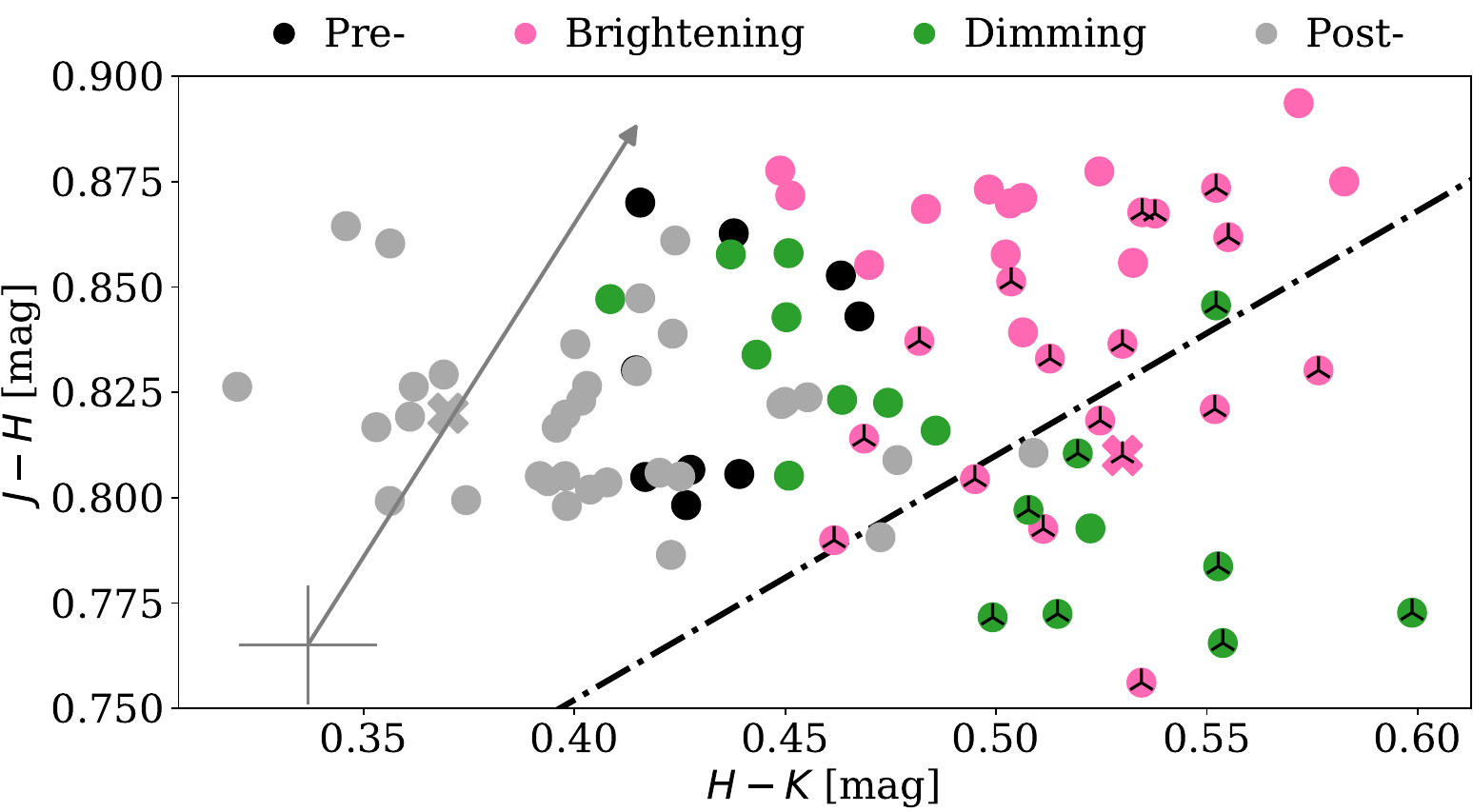}
  \caption{$J-H$ vs $H-K$ color-color diagram. The marker colors indicate the stages of the outburst as labeled in Fig.~\ref{fig:cmd}. The circle and X symbols are the REM photometry and the X-shooter synthetic photometry, respectively. The black tri-markers indicate the colors when \exl{} was in outburst, i.e.\ when $g\leq12.5$\,mag. In the bottom left, the light gray cross shows the median uncertainties of both photometric colors, and the arrow is the $A_V$ = 1.1\,mag extinction vector following the extinction curve of \citet{Cardelli1989_ApJ345245C}. The dash–dotted line is the locus of unreddened T~Tauri stars \citep{Meyer1997_AJ114288M}.\label{fig:ccd}}
\end{figure}

Our long-term multi-filter photometric monitoring can be used to construct color-magnitude and color-color diagrams, which we can then use to test whether the brightening of \exl{} is caused by a decrease in the extinction or by an increase of the luminosity of the system.
In Fig.~\ref{fig:cmd} we present the color-magnitude diagrams using the seven photometric filters, and in Fig.~\ref{fig:ccd} the color-color diagram constructed with the three near-infrared filters.
In both diagrams, the symbols are color-coded depending on the stage of the light curve as shown in the bottom of Fig.~\ref{fig:lightcurve}.

A visual comparison between our photometric data and the extinction vectors shows that, as reported by \citet{Kospal2022_RNAAS652K}, the burst did not follow the extinction vector.
For a quick numerical check, we carried out a linear-fit to each color-magnitude diagram and compared the fitted slope with the extinction vector from the extinction curve of \citet{Cardelli1989_ApJ345245C} with $R_V$=3.1.
We found that, with the exception of the $r-i$ color, all colors follow a different path than the extinction, which suggests that the brightening was not due to a decrease in the line-of-sight extinction.
In the $g$ vs. $g-r$ diagram, a slight ``knee'' is seen at $g\approx$ 12.5\,mag and it is at this magnitude that \exl{} was unequivocally in burst, while data points when \exl{} was fainter than this magnitude mean that the source was in quiescence.
The $i-z$ color panel shows an interesting behavior where the brightening and the dimming stages follow different paths, with \exl{} getting slightly redder than when in quiescence while the dimming getting bluer.

The near-infrared color-color diagram (Fig.~\ref{fig:ccd}) also shows how the colors of \exl{} have changed during the burst.
The figure shows that an $A_V = 1.1$ suggests that \exl{} is a slightly extincted T~Tauri star and, more importantly, that a change in extinction cannot explain the changes in the near-infrared colors seen during our monitoring of the burst.
The changes in color during its 2022 burst are both significantly smaller and with a different direction than those seen in other EXor-type sources during their (out)bursts, including \exl{} itself during its powerful 2008 outburst \citep{Lorenzetti2009_ApJ6931056L,Lorenzetti2012_ApJ749188L}.
Indeed, while the other EXors shift to bluer $J-H$ and $H-K$ colors, during its 2022 burst \exl{} became bluer in $J-H$ and redder in $H-K$ (see also the two rightmost panels of Fig.~\ref{fig:cmd}).

\section{Results and Analysis: Spectroscopy}\label{sec:spec}
The three spectra obtained for \exl{} with X-shooter are presented in Fig.~\ref{fig:plot_all}.
During the peak of the burst, the level of the continuum is higher across the full wavelength coverage, and the overall shape of the spectrum is much flatter than for the two epochs in quiescence.
The difference in flux levels between the two quiescent snapshots and the peak of the burst is, as expected, larger at shorter wavelengths, suggesting the appearance of an extra hot continuum during the burst.
The Balmer jump is not seen in the burst spectrum, while it is clearly seen in the two quiescent spectra.
Between the two quiescent measurements, the 2022 July spectrum has some UV excess with respect to the 2010 one.
This excess can be interpreted as a slightly elevated \macc{} leftover from the 2022 burst, as an intrinsic variability that occurs even during quiescence, or as changes related to the 7-day modulation found by \citet{Kospal2014_AA561A61K}.

\exl{} is an M0-type star, thus its spectra feature several photospheric absorption lines \citep{Herbig2001_PASP1131547H}.
We removed the photospheric contribution from our three X-shooter spectra by using archival observations with same instrument of an M0 pre-main sequence (or Class~III) template star, TYC~7760-283-1 (085.C-0238, PI: J.M. Alcalá).
To do this subtraction, we carried out an iterative fitting of the stellar template to the radial velocity and veiling of \exl{}.
We split our continuum-normalized spectra into 5\,nm chunks, and varied these two parameters until we minimized the difference between our target and the template star.
The best fits of some spectral chunks resulted in radial velocities that were significantly different from the median, thus we shifted the complete photospheric spectrum to the median radial velocity of each epoch, and fitted only the veiling for each spectral chunk.
Finally, we subtracted the absorption component of our best-fitted template spectra from our \exl{} measurements.

\subsection{Emission lines}
\begin{figure*}
  \centering
  \includegraphics[width=\linewidth]{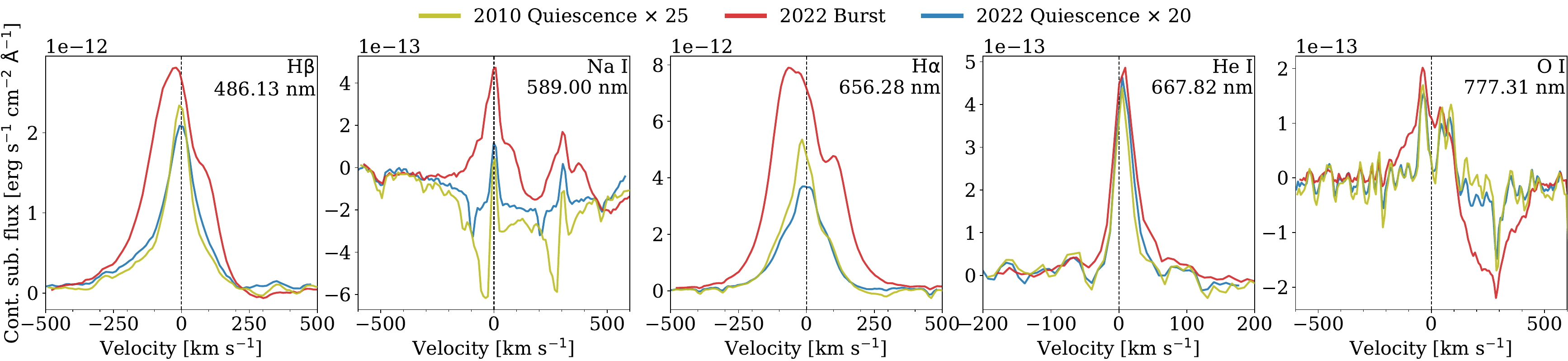}
  \caption{The continuum-subtracted spectra of five emission lines found in the three epochs of \exl{} observations that we used to calculate the \lacc{}. Note to ease the visual comparison between epochs, the 2010 and 2022 spectra have been scaled up by factors of 25 and 20, respectively. Below the identified of each line is the rest wavelength for each transition, with the exception of the O~I triplet where the wavelength is in between the three transitions. The yellow, red and blue lines represent the 2010 quiescent, 2022 bursting, and 2022 post-burst quiescent spectra taken with X-shooter. The \textit{x}-axes are the heliocentric-corrected velocities calculated with respect to the rest wavelength, and the vertical dashed line indicates the zero-velocity. A visual inspection of these spectral lines shows how the burst strengthened and broadened the lines, and, in the case of the O~I triplet, how the blending worsened as a result of this. \label{fig:alcala_feat}}
\end{figure*}

As previous works have shown, the \exl{} spectrum (Fig.~\ref{fig:plot_all}) is crowded with lines \citep{Herbig2001_PASP1131547H,Herbig2007_AJ1332679H,Kospal2011_ApJ73672K,SiciliaAguilar2012_AA544A93S,Alcala2017_AA600A20A,Rigliaco2020_AA641A33R,CampbellWhite2021_MNRAS5073331C}.
Similar to these previous studies, our observations show that the \exl{} spectrum has multiple hydrogen emission lines of the Brackett, Balmer, and Paschen series.
In addition, it exhibits multiple emission lines of neutral atoms (e.g.\ Na, Ca, Fe, Ti, K, Ni, Mn, Cr, Co, and V) and of ionized species (e.g.\ Ti, Sc, He, Ca, Cr, Fe).
We used the National Institute of Standards and Technology (NIST) Atomic Spectra Database\footnote{\url{https://physics.nist.gov/PhysRefData/ASD/lines_form.html}} to identify the emission lines found for each of these atoms; however, an in-depth presentation of the lines will be prepared for a follow-up publication.
Unsurprisingly, the spectrum taken close to the peak of the burst shows the largest number of lines, which are also stronger and broader than the pre- and post-burst spectra.
In Fig.~\ref{fig:alcala_feat} we show a few of the lines we identified.
A comparison between the three epochs is discussed in Sect.~\ref{ss:linevariability}.

\subsection{Accretion luminosity}\label{ss:acclum}
Our first method to calculate the \lacc{} and \macc{} is by using the empirical relationships between line luminosity and accretion luminosity as determined by \citet{Alcala2017_AA600A20A} for 38 emission lines which include several transitions for the Balmer, Paschen, and Brackett hydrogen series, plus some He~I, He~II, Ca~II, Na~I and O~I lines. 
For the first step of this process, we corrected the three photosphere-subtracted spectra for interstellar extinction using the extinction curve of \citet{Cardelli1989_ApJ345245C} with $R_V$ = 3.1, and assuming $A_V$ = 1.1\,mag, as estimated by \citet{Alcala2017_AA600A20A}.
Next, we generated a sub-spectrum per accretion tracer, using $\pm$2000\,km\,s$^{-1}$ as limits for each window.
We estimated the continuum of each spectral window by fitting a straight line to the 2-sigma clipped spectral window.
After subtracting the continuum, we computed the line flux ($F_\mathrm{line}$) by integrating all the emission inside a narrow spectral window tailored for each line.
Finally, we used this line flux to calculate the line luminosity as $L_\mathrm{line} = 4 \pi d^2 F_\mathrm{line}$ and the accretion luminosity as $\log{\lacc{}} = a \log{L_\mathrm{line}} + b$, where $d$ is the distance to the source, and $a$ and $b$ are parameters empirically obtained by \citet{Alcala2017_AA600A20A}.
In Fig.~\ref{fig:alcala_feat} we show five of the lines we used for this computation.
In Table~\ref{table:macc} we present our values of \lacc{} and \macc{} for each line in each epoch, and in Tables~\ref{table:lflx_llum_2010}, \ref{table:lflx_llum_202203} and \ref{table:lflx_llum_202207} we present the line fluxes and line luminosities.
As mentioned earlier, our main goal is to study how the \lacc{} and \macc{} have changed during the outburst, therefore, we only included the uncertainties of the line fluxes, and of the $a$ and $b$ coefficients from \citet{Alcala2017_AA600A20A} because considering the uncertainties of the distance would only increase the size of the error bars without providing information for our analysis.
However, a comparison of \exl{} with other young stars should take it into account.

\begin{table*}
  \caption{Accretion luminosities (\lacc{}) and mass accretion rates (\macc{}) for the three X-SHOOTER epochs. The \lacc{} values were calculated using the two methods explained in Sect.~\ref{ss:acclum}. The \macc{} values shown here were calculated using the stellar radius and mass from \citet{Sipos2009_AA507881S}, i.e.\ $R_\star$~=~1.6\,R$_\odot$ and $M_\star$~=~0.6\,M$_\odot$.\label{table:macc}}
\centering
\begin{tabular}{lrccccccccc}
\hline\hline
& & \multicolumn{2}{c}{2010 May 04} & & \multicolumn{2}{c}{2022 Mar 27} & & \multicolumn{2}{c}{2022 Jul 29}\\
\cline{3-4} \cline{6-7} \cline{9-10}
Line & \multicolumn{1}{c}{$\lambda$} & $L_\mathrm{acc}$ & \macc{} $\times$ 10$^{-9}$  & & $L_\mathrm{acc}$ & \macc{} $\times$ 10$^{-7}$  & & $L_\mathrm{acc}$ & \macc{} $\times$ 10$^{-9}$\\
& \multicolumn{1}{c}{[nm]} & [L$_\odot$] & [M$_\odot$\,yr$^{-1}$] & & [L$_\odot$] & [M$_\odot$\,yr$^{-1}$] & & [L$_\odot$] & [M$_\odot$\,yr$^{-1}$]\\
\hline
H15         & 371.20  & 0.04$\pm$0.03 & 4.69$\pm$3.17   &  & 1.03$\pm$0.63  & 1.10$\pm$0.67  &  & 0.04$\pm$0.03 & 3.93$\pm$2.68 \\
H14         & 372.19  & 0.05$\pm$0.03 & 4.98$\pm$3.16   &  & 2.20$\pm$1.21  & 2.34$\pm$1.28  &  & 0.04$\pm$0.03 & 4.77$\pm$3.03 \\
H13         & 373.44  & 0.06$\pm$0.03 & 6.09$\pm$3.68   &  & 2.31$\pm$1.21  & 2.45$\pm$1.28  &  & 0.06$\pm$0.03 & 5.95$\pm$3.60 \\
H12         & 375.02  & 0.07$\pm$0.03 & 7.21$\pm$3.61   &  & 1.71$\pm$0.78  & 1.81$\pm$0.82  &  & 0.05$\pm$0.03 & 5.44$\pm$2.75 \\
H11         & 377.06  & 0.08$\pm$0.04 & 8.86$\pm$4.37   &  & 2.75$\pm$1.23  & 2.92$\pm$1.31  &  & 0.09$\pm$0.04 & 9.59$\pm$4.72 \\
H10         & 379.79  & 0.08$\pm$0.04 & 8.37$\pm$4.68   &  & 3.05$\pm$1.47  & 3.23$\pm$1.57  &  & 0.10$\pm$0.06 & 11.02$\pm$6.09 \\
H9          & 383.54  & 0.10$\pm$0.06 & 11.14$\pm$6.13  &  & 1.14$\pm$0.57  & 1.21$\pm$0.61  &  & 0.13$\pm$0.07 & 13.97$\pm$7.61 \\
H8          & 388.90  & 0.13$\pm$0.07 & 13.84$\pm$7.20  &  & 4.54$\pm$2.05  & 4.82$\pm$2.18  &  & 0.18$\pm$0.09 & 19.40$\pm$9.94 \\
Ca II K     & 393.37  & 0.09$\pm$0.05 & 9.74$\pm$5.08   &  & 17.02$\pm$7.31 & 18.07$\pm$7.76 &  & 0.14$\pm$0.07 & 15.02$\pm$7.68 \\
Ca II H     & 396.85  & 0.09$\pm$0.04 & 9.31$\pm$4.10   &  & 8.62$\pm$3.32  & 9.15$\pm$3.52  &  & 0.15$\pm$0.07 & 16.00$\pm$6.92 \\
H$\epsilon$ & 397.01  & 0.10$\pm$0.05 & 10.29$\pm$5.40  &  & 4.22$\pm$1.91  & 4.48$\pm$2.02  &  & 0.11$\pm$0.06 & 12.08$\pm$6.30 \\
He I        & 402.62  & 0.09$\pm$0.06 & 9.28$\pm$6.07   &  & 3.79$\pm$2.17  & 4.02$\pm$2.31  &  & 0.09$\pm$0.06 & 9.62$\pm$6.29 \\
H$\delta$   & 410.17  & 0.12$\pm$0.06 & 12.71$\pm$6.55  &  & 4.79$\pm$2.14  & 5.09$\pm$2.28  &  & 0.19$\pm$0.09 & 19.84$\pm$10.04 \\
H$\gamma$   & 434.05  & 0.11$\pm$0.05 & 11.69$\pm$5.29  &  & 5.47$\pm$2.24  & 5.80$\pm$2.38  &  & 0.17$\pm$0.08 & 18.48$\pm$8.25 \\
He I        & 447.15  & 0.10$\pm$0.06 & 10.71$\pm$6.86  &  & 2.67$\pm$1.53  & 2.84$\pm$1.63  &  & 0.12$\pm$0.08 & 13.10$\pm$8.33 \\
He II       & 468.58  & 0.11$\pm$0.11 & 12.15$\pm$11.26 &  & 5.07$\pm$4.24  & 5.38$\pm$4.50  &  & 0.13$\pm$0.12 & 14.24$\pm$13.13 \\
He I        & 471.31  & 0.04$\pm$0.06 & 4.35$\pm$6.16   &  & 0.62$\pm$0.78  & 0.66$\pm$0.83  &  & 0.05$\pm$0.07 & 5.00$\pm$7.02 \\
H$\beta$    & 486.13  & 0.08$\pm$0.04 & 8.54$\pm$4.05   &  & 6.55$\pm$2.59  & 6.96$\pm$2.75  &  & 0.12$\pm$0.06 & 12.66$\pm$5.89 \\
He I Fe I   & 492.19  & 0.08$\pm$0.05 & 8.53$\pm$5.80   &  & 6.74$\pm$4.00  & 7.16$\pm$4.24  &  & 0.08$\pm$0.06 & 9.01$\pm$6.12 \\
He I        & 501.57  & 0.05$\pm$0.03 & 4.87$\pm$3.47   &  & 2.76$\pm$1.72  & 2.93$\pm$1.82  &  & 0.05$\pm$0.04 & 5.84$\pm$4.13 \\
He I        & 587.56  & 0.08$\pm$0.05 & 8.41$\pm$5.18   &  & 3.95$\pm$2.14  & 4.19$\pm$2.27  &  & 0.10$\pm$0.06 & 10.60$\pm$6.48 \\
Na I        & 589.00  & 0.02$\pm$0.02 & 1.96$\pm$2.08   &  & 3.60$\pm$3.24  & 3.82$\pm$3.44  &  & 0.03$\pm$0.03 & 3.48$\pm$3.63 \\
Na I        & 589.59  & 0.02$\pm$0.02 & 1.61$\pm$1.80   &  & 2.02$\pm$1.95  & 2.14$\pm$2.07  &  & 0.03$\pm$0.03 & 3.36$\pm$3.69 \\
H$\alpha$   & 656.28  & 0.03$\pm$0.02 & 3.32$\pm$1.83   &  & 3.34$\pm$1.52  & 3.55$\pm$1.61  &  & 0.03$\pm$0.02 & 3.55$\pm$1.95 \\
He I        & 667.82  & 0.09$\pm$0.09 & 9.83$\pm$9.72   &  & 8.52$\pm$7.39  & 9.05$\pm$7.84  &  & 0.13$\pm$0.13 & 14.29$\pm$13.97 \\
He I        & 706.52  & 0.07$\pm$0.06 & 7.27$\pm$6.29   &  & 4.13$\pm$3.17  & 4.39$\pm$3.36  &  & 0.08$\pm$0.07 & 8.19$\pm$7.06 \\
O I         & 777.31  & 0.05$\pm$0.08 & 5.40$\pm$8.05   &  & 6.42$\pm$8.30  & 6.82$\pm$8.81  &  & 0.06$\pm$0.09 & 6.76$\pm$9.99 \\
O I         & 844.64  & 0.01$\pm$0.03 & 1.42$\pm$2.81   &  & 5.14$\pm$8.18  & 5.45$\pm$8.68  &  & 0.02$\pm$0.04 & 2.20$\pm$4.28 \\
Ca II       & 849.80  & 0.05$\pm$0.04 & 5.71$\pm$4.60   &  & 13.39$\pm$9.23 & 14.22$\pm$9.80 &  & 0.06$\pm$0.05 & 6.32$\pm$5.07 \\
Ca II       & 854.21  & 0.05$\pm$0.04 & 5.04$\pm$4.31   &  & 10.34$\pm$7.21 & 10.97$\pm$7.65 &  & 0.06$\pm$0.05 & 6.19$\pm$5.25 \\
Ca II       & 866.21  & 0.05$\pm$0.04 & 4.79$\pm$4.20   &  & 7.61$\pm$5.49  & 8.07$\pm$5.83  &  & 0.05$\pm$0.05 & 5.50$\pm$4.80 \\
Pa10        & 901.49  & 0.01$\pm$0.01 & 0.83$\pm$1.34   &  & 0.81$\pm$1.11  & 0.86$\pm$1.18  &  & 0.01$\pm$0.01 & 0.65$\pm$1.06 \\
Pa9         & 922.90  & 0.02$\pm$0.02 & 1.99$\pm$2.59   &  & 1.78$\pm$2.01  & 1.89$\pm$2.14  &  & 0.01$\pm$0.02 & 1.37$\pm$1.81 \\
Pa8         & 954.60  & 0.03$\pm$0.06 & 3.70$\pm$6.66   &  & 1.79$\pm$2.78  & 1.90$\pm$2.95  &  & 0.03$\pm$0.06 & 3.36$\pm$6.07 \\
Pa$\delta$  & 1004.94 & 0.04$\pm$0.05 & 3.94$\pm$5.21   &  & 3.60$\pm$4.06  & 3.82$\pm$4.31  &  & 0.03$\pm$0.05 & 3.65$\pm$4.84 \\
Pa$\gamma$  & 1093.81 & 0.03$\pm$0.03 & 3.39$\pm$2.85   &  & 3.34$\pm$2.37  & 3.55$\pm$2.51  &  & 0.03$\pm$0.02 & 3.05$\pm$2.57 \\
Pa$\beta$   & 1281.81 & 0.02$\pm$0.02 & 2.26$\pm$2.33   &  & 2.14$\pm$1.85  & 2.27$\pm$1.97  &  & 0.02$\pm$0.02 & 2.39$\pm$2.46 \\
Br$\gamma$  & 2166.12 & 0.01$\pm$0.02 & 1.46$\pm$2.39   &  & 1.41$\pm$1.97  & 1.50$\pm$2.09  &  & 0.01$\pm$0.02 & 1.04$\pm$1.72 \\
\hline
\multicolumn{2}{c}{Slab model} & 0.16 & 17.20 & & 1.48 & 1.57 & & 0.29 & 31.00\\
\hline
\end{tabular}
\end{table*}

For the second method, we fitted slab models to the shorter wavelengths of the spectra ($<$1000\,nm) without subtracting the photosphere and without correcting for extinction.
The method is described in \citet{Manara2013_AA558A114M} and has been used in the study of accretion in young stellar objects \citep[e.g.][]{Manara2016_AA585A136M,Alcala2017_AA600A20A,Claes2022_AA664L7C}.
It consists of finding the best $\chi^2$ fit between the observed spectrum, and the sum of a photospheric template and a slab model, both reddened with an extinction $A_V$.
The slab model represents the excess emission due to accretion.
The best fit model is then used to estimate the accretion luminosity, \lacc{}.

\begin{figure*}
  \centering
  \includegraphics[width=\linewidth]{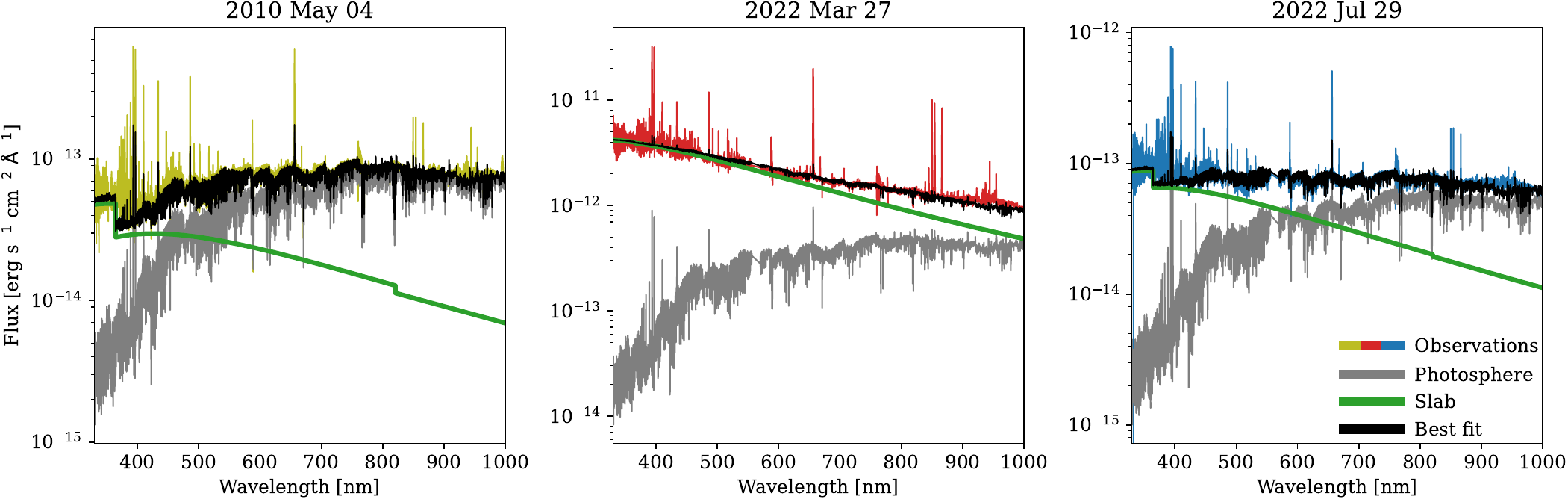}
  \caption{Best-fitting slab models for the three epochs. Each panel contains the dereddened spectrum (each epoch represented by a different color), the photosphere template (gray), the best-fit slab (green), and the total combination of the slab and photosphere (black) used to calculate the $\chi^2$ during the fitting process.\label{fig:slab}}
\end{figure*}

We ran the fitting procedure for our three X-shooter spectra with a fixed spectral type (M0) and extinction ($A_V$ = 1.1).
The 2010 spectrum has been fitted by \citet{Alcala2017_AA600A20A} using the slab model, we re-ran this analysis due to the difference in the slit-loss correction (see Sect.~\ref{ss:xshooter}) and the fixed parameters.
The fitting procedure also uses the previously mentioned distance to the object to scale the flux.
In Fig.~\ref{fig:slab} we show a visual representation of our best-fitted results, where we show the observed spectra, the photospheric template, the emission of the accretion slab, and the total model for the three epochs.
Using the best-fit values we calculated \lacc{} values of 0.162\,L$_\odot$, 1.48\,L$_\odot$ and 0.292\,L$_\odot$ for the 2010 May, 2022 March and 2022 July epochs, respectively.

\begin{figure*}
 \centering
 \includegraphics[width=\textwidth]{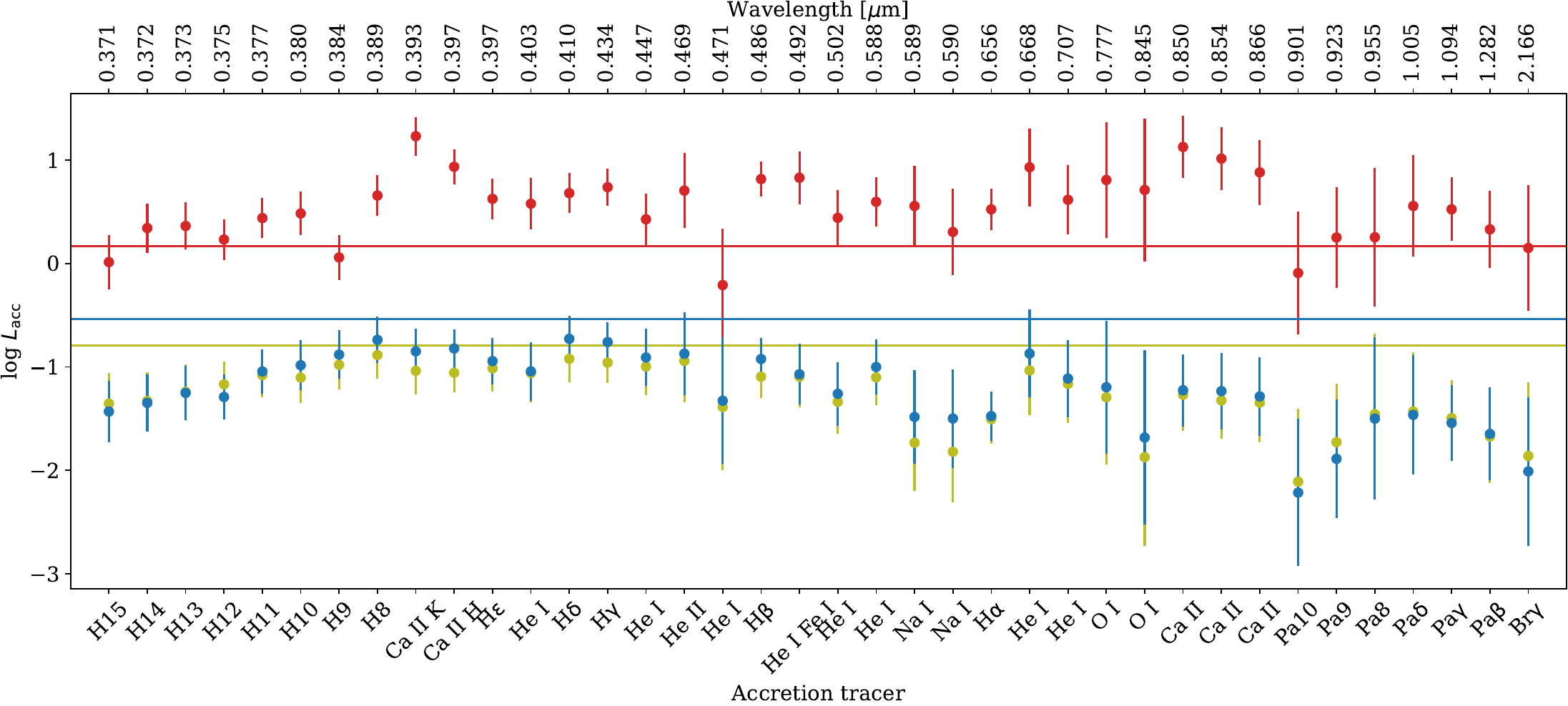}
 \caption{Accretion luminosity (\lacc{}) calculated for each line per epoch. Each epoch is represented by a different color as shown earlier: yellow for the 2010 May 04 epoch, red for 2022 March 27, and blue for 2022 July 29. The horizontal lines are the values obtained from the slab model fitting per epoch.\label{fig:lacc}}
\end{figure*}

In Fig.\ref{fig:lacc} we show a comparison of the \lacc{} calculated from each line per epoch with the two different methods.
A comparison between these is discussed in Sect.~\ref{ss:lacc}.

\subsection{Mass accretion rate}\label{ss:macc_lines_slab}
We calculated the mass accretion rate taking the \lacc{} calculated for each emission line and from the slab modeling, and used it as input for the relationship between \lacc{} and \macc{} during magnetospheric accretion \citep{Gullbring1998_ApJ492323G}:
\begin{equation}
  \dot{M}_\mathrm{acc} = \left(1-\frac{R_\star}{R_\mathrm{in}}\right)^{-1} \frac{L_\mathrm{acc}R_\star}{\mathrm{G}M_\star} \approx \frac{1.25\ L_\mathrm{acc}\ R_\star}{\mathrm{G}\ M_\star},
\end{equation}
where G is the gravitational constant, $R_\star$ and $M_\star$ are the stellar radius and mass, respectively, and $R_\mathrm{in}$ is the inner disk radius, for which we have assumed $R_\mathrm{in} = 5 R_\star$ \citep{Hartmann2016_ARAA54135H}.
For these two latter parameters, we first used the values determined by \citet{Sipos2009_AA507881S}: $R_\star$~=~1.6\,R$_\odot$ and $M_\star$~=~0.6\,M$_\odot$.
Our resulting \macc{} values are shown in Table~\ref{table:macc}.

The slab fitting procedure also finds the best-fit for the stellar luminosity per spectrum from which a stellar radius is calculated.
It also finds a stellar mass by interpolating the best-fit stellar luminosity and effective temperature on different stellar tracks for young objects; see \citet{Manara2013_AA558A114M} for details.
This results in each epoch having slightly different stellar radius and mass, and each epoch being different from the ones calculated by \citet{Sipos2009_AA507881S}.
Using the \citet{Baraffe2015_AA577A42B} tracks we obtained \macc{} values of \sci{2.34}{-8}\,M$_\odot$\,yr$^{-1}$, \sci{1.62}{-7}\,M$_\odot$\,yr$^{-1}$ and \sci{3.58}{-8}\,M$_\odot$\,yr$^{-1}$ for the 2010 May, 2022 March and 2022 July epochs, respectively.
Alternatively, the values obtained by using the \citet{Siess2000_AA358593S} tracks are \sci{2.22}{-8}\,M$_\odot$\,yr$^{-1}$, \sci{1.57}{-7}\,M$_\odot$\,yr$^{-1}$ and \sci{3.41}{-8}\,M$_\odot$\,yr$^{-1}$, respectively.
We can only directly compare our results from both methods if we use the same stellar parameters. Thus, by taking the stellar parameters from \citet{Sipos2009_AA507881S}, we find \macc{} values of \sci{1.72}{-8}\,M$_\odot$\,yr$^{-1}$, \sci{1.57}{-7}\,M$_\odot$\,yr$^{-1}$ and \sci{3.10}{-7}\,M$_\odot$\,yr$^{-1}$ for the 2010 May, 2022 March and 2022 July epochs, respectively.
Below we compare these values with the ones obtained using the emission lines.

\subsection{Evolution of the mass accretion rate during the outburst}\label{ss:acc_curve}
\begin{figure*}
  \centering
  \includegraphics[width=\linewidth]{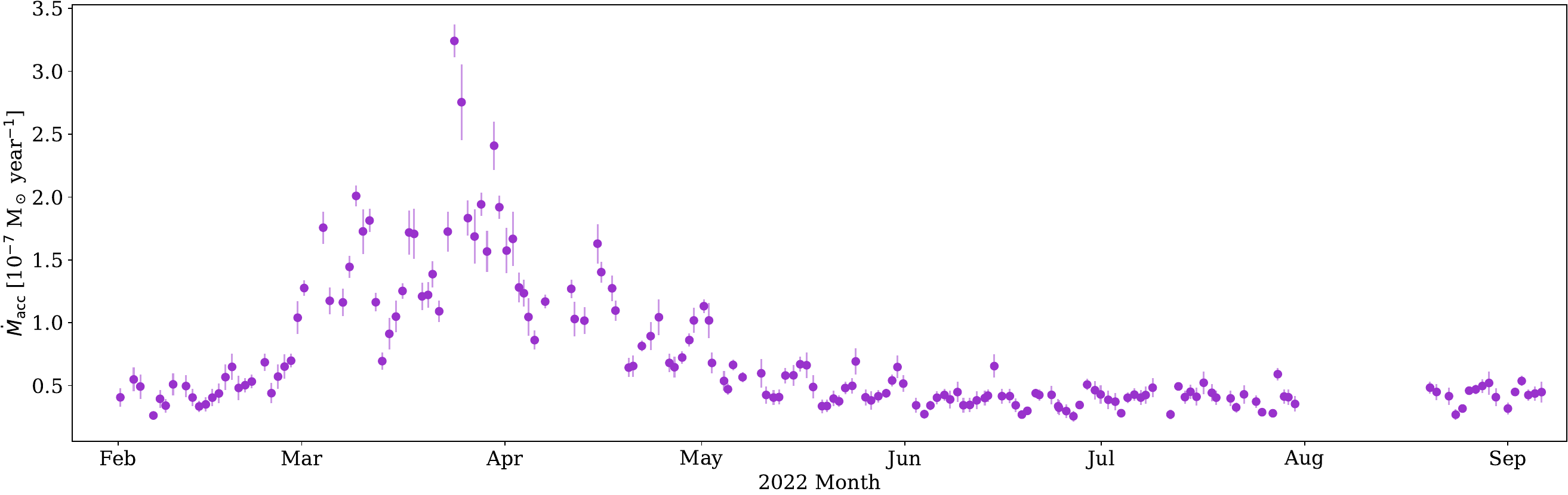}
  \caption{Mass accretion rate as a function of time, see Sect.\ \ref{ss:acc_curve}. The peak of the outburst reached a \macc{} value 7 times higher than the quiescent level, and the total mass integrated during the outburst was \sci{2.9}{-8}\,M$_\odot$, or 0.8 Lunar masses.\label{fig:macc_curve}}
\end{figure*}

The previous two analyses provided estimates of the \macc{} at specific times, which is a strong constraint for our goals of learning how the \macc{} evolved through the outburst, and what the total mass accreted by \exl{} was during this event.
To follow the evolution of \macc{} along the burst in a more detailed way, we decided to use the results from our slab modeling of the 2022 July 29 (i.e.\ post-burst) epoch as anchor to translate the light curves from the photometric monitoring into an \emph{accretion curve}.
The procedure is similar to that of the spectroscopic fitting but, instead of fitting to a spectrum, we calculated synthetic photometry of a large grid of models (i.e.\ the reddened sum of photosphere and slab) at the three optical filters ($g'r'i'$) and then found the best fitting model from the grid for each photometric epoch.
The slabs are defined with three free parameters: with different $T_\mathrm{slab}$, $n_e$, and $\tau$.
This procedure has fewer points to carry out the fitting, thus our results are degenerate, i.e.\ slabs with different properties can fit the photometric points within the error bars.
However, we found that the slab models which give the best match (lowest $\chi^2$) result in \lacc{} values within a relatively narrow range.
Thus, while the parameters of the best-fitting models are degenerate, their \lacc{} values are not. 
We took these best fitted \lacc{} estimates, averaged them to obtain an approximation of the \lacc{} per photometric epoch, and derived \macc{} from them.
We re-ran our calculations using the 2022 March 27 measurement (i.e.\ almost at the peak of the burst) as the anchor, however, the differences with respect to our chosen epoch are of less than 8\%.
Our resulting \emph{accretion curve} is shown in Fig.~\ref{fig:macc_curve}, and we discuss its significance in the following Section.

\section{Discussion}\label{sec:disc}
We have shown that the brightening experienced by \exl{} in 2022 was caused by an increase in the mass accretion rate and not by a change in the line-of-sight extinction.
\exl{} has gone through more than a dozen bursts that have reached magnitudes brighter than V$\sim$11\,mag, such as the one presented in this work.
Outbursts as powerful as the three previously experienced by \exl{} have been shown to cause important changes in the mineralogy of the disk by the creation of silicate crystals \citep{Abraham2009_Natur459224A} and molecular abundances \citep{Banzatti2012_ApJ74590B,Banzatti2015_ApJ798L16B}.
The impact that bursts can have on a system, either individually or collectively, is still understudied.

\subsection{Line variability}\label{ss:linevariability}
Here we present a brief qualitative discussion of the features found in the emission lines of the three X-shooter spectra.
A more detailed analysis of this is reserved for a future publication.
In Fig.~\ref{fig:alcala_feat} we show a sub-set of continuum-subtracted emission lines that we will use to showcase the differences and similarities between the three epochs.
One of our snapshots was taken within a few days from the peak of the burst (red lines in Fig.~\ref{fig:alcala_feat}), which is when we expect higher gas temperatures due to the increased accretion rate, thus we expect this spectrum to show the largest differences with respect to the other two (blue and yellow lines in Fig.~\ref{fig:alcala_feat}).

We found that during the peak of the burst, some lines become stronger and broader (e.g.\ Na~I, O~I, H$\alpha$), as expected from the analysis of the 2008 outburst of \exl{} \citep{Aspin2010_ApJ719L50A,Goto2011_ApJ7285G,Kospal2011_ApJ73672K,SiciliaAguilar2012_AA544A93S,SiciliaAguilar2015_AA580A82S,Banzatti2015_ApJ798L16B}.
Some lines showed higher levels of asymmetry (e.g.\ H$\alpha$ and H$\beta$), broad components reaching velocities in the order of a few 100\,km\,s$^{-1}$ (e.g.\ He~I), and strong red-shifted absorption (e.g.\ O~I triplet).
As shown in the previous works, the blending between lines increased when \exl{} was in burst.
In the case of the O~I triplet, we can see that during quiescence, the 777.4\,nm and 777.5\,nm lines are blended while the 777.1\,nm line is not, while in burst the three lines are blended with each other.

Our last snapshot was taken after \exl{} went back to its pre-outburst levels of \macc{}, thus, we expected its spectrum to be similar to the one taken in 2010.
However, we found that some lines have features that are in between the outburst and the quiescent spectra.
For example, the Na~I blue-shifted absorption has not returned to its pre-outburst shape and its emission components remain broader.
In the case of O~I triplet, the post-outburst emission lines are the same as the pre-outburst ones but the red-shifted absorption feature seen during outburst is still seen albeit weaker.
These differences indicate that even though the \macc{} has gone back to normal, the physical properties (i.e.\ temperature and density) of the gas have not entirely done so.
Therefore, any evolution caused by these outburst-driven changes could still be occurring months after the outburst ended.

Finally, we point out that the variations mentioned here should be taken with caution because the spectral lines of \exl{} have shown variations that are not related to the accretion outburst \citep{SiciliaAguilar2012_AA544A93S,SiciliaAguilar2015_AA580A82S,CampbellWhite2021_MNRAS5073331C}, and, thus, we expect that the differences discussed here are caused by both the accretion burst and the rotational effects.

\subsection{Searching for winds and jets with [O~I]}
The [O~I] line at 630\,nm is often used to study outflows, including those emanating from young T~Tauri stars.
There is a correlation between the kinematic properties obtained from analyzing the line and the \macc{} of their ejecting stars \citep[e.g.][]{Nisini2018_AA609A87N,Banzatti2019_ApJ87076B,Gangi2020_AA643A32G,Gangi2022_AA667A124G}.
Since our spectral data set covers this line and we know that there was a temporary increase in the accretion, the 2022 burst of \exl{} is a great opportunity to study how the line changed due to the variations in the \macc{}.
Here we present a short qualitative description of these changes, as a more in-depth analysis is being prepared for a follow-up publication.

\begin{figure*}
  \centering
  \includegraphics[width=\linewidth]{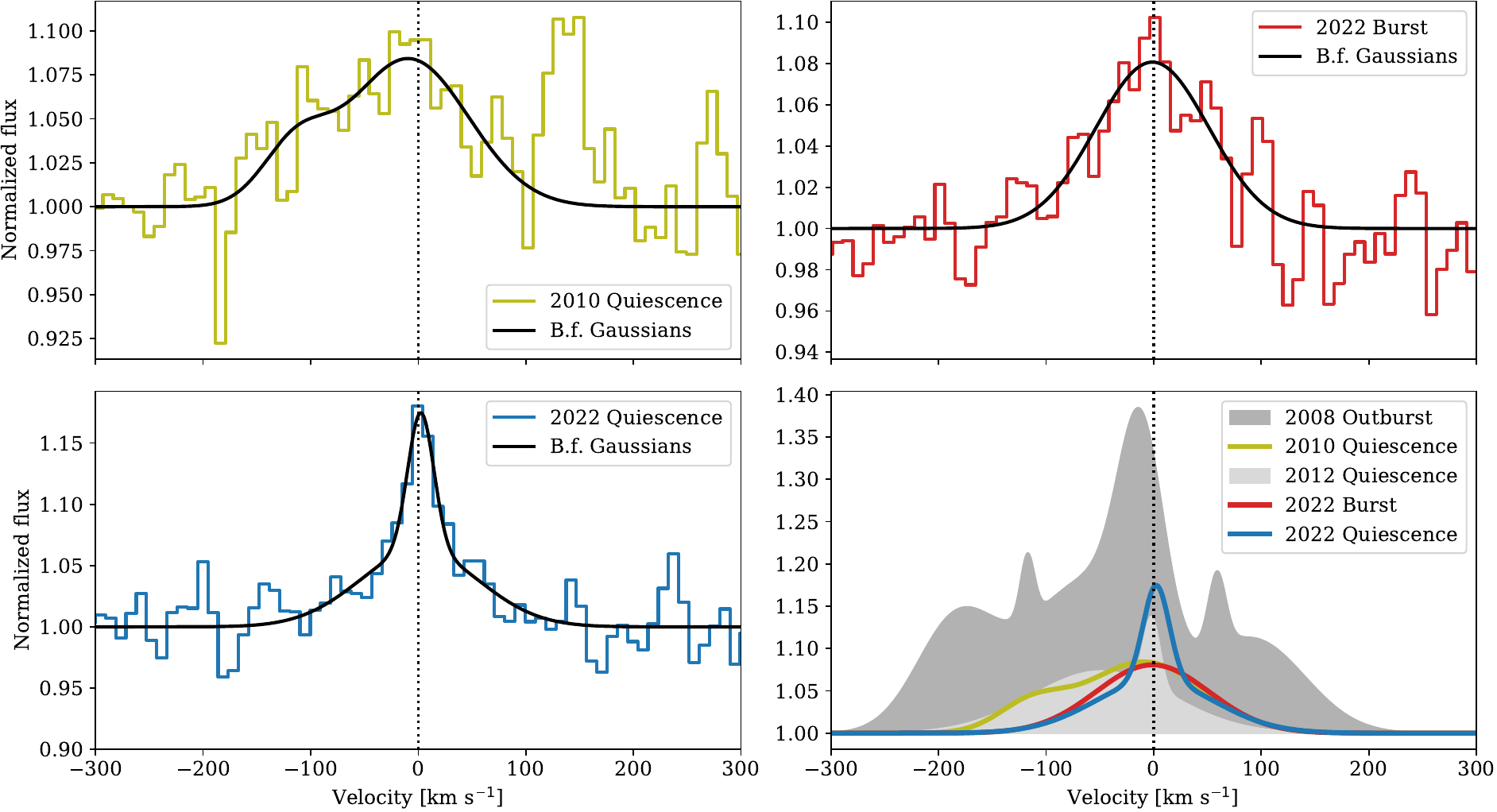}
  \caption{Evolution of the [O~I] line due through the 2022 burst and an comparison with the line profiles seen during and after the 2008 outburst. The top panels and the bottom left panel show the [O~I] line profile from the three X-shooter spectra analyzed in this work as a colored line, and the best-fit Gaussians in a black line. In the bottom right panel we use colored lines to show the best-fit Gaussians, and shaded gray areas to show the best-fit Gaussians by \citet{Banzatti2019_ApJ87076B} calculated on their high-resolution spectra taken during the 2008 outburst and in quiescence in 2012.\label{fig:oi}}
\end{figure*}

We extracted a narrow spectral window from the spectrum of each epoch, which we show in Fig.~\ref{fig:oi} plotted with respect to the heliocentric-corrected velocities.
In the case of moderate-resolution spectra of T~Tauri stars, the line profile of [O~I] is often separated into two components, one with low-velocity (i.e.\ within a few km\,s$^{-1}$ from the systemic velocity of the star) and one with high-velocity ($\geq$40\,km\,s$^{-1}$ away from the systemic).
Therefore, we fitted each spectral window using two Gaussian functions, and the best fits are also shown in Fig.~\ref{fig:oi}.
The 2010 spectrum shows a strong feature in emission at $\sim$130\,km\,s$^{-1}$ but we do not consider this as real redshifted [O~I] emission and we do not take it into consideration in this discussion.
\citet{Banzatti2019_ApJ87076B} carried out a study dedicated to this line using high spectral resolution near-infrared spectra of 64 T~Tauri stars.
Their sample included two spectra of \exl{}, one taken during its 2008 outburst and another in its post-outburst stage in 2012r, and they fitted multiple Gaussian functions to each spectrum.
In the bottom right panel of Fig.~\ref{fig:oi}, we show our best-fit Gaussians and the two from \citet{Banzatti2019_ApJ87076B}.

The best fit of the 2010 spectrum is composed by two blueshifted broad components.
The one with the highest velocity (-115\,km\,s$^{-1}$) has a comparable shape to the one found by \citet{Banzatti2019_ApJ87076B} in the 2012 spectrum but we did not recover the narrower low-velocity component seen by them.
During the 2022 burst, [O~I] is only characterized by a broad low-velocity component.
Indeed, fitting a single Gaussian produces an equally good result.
Finally, the best-fit found for the post-burst spectrum is composed by a two low-velocity components, one broad and one narrow.
The centroid velocities for all the low-velocity components are comparable with the spectral resolution of our observations, and, thus, it is not clear whether they are blue- or redshifted.

The high-velocity component is used as a tracer of high-velocity jet emission and the low-velocity as a tracer of slow winds.
Therefore, the 2008 outburst (dark gray shaded area) generated strong blue- and redshifted jet emission and strong blueshifted slow winds.
By 2010 (yellow line) and 2012 (light gray area), the redshifted emission is not seen but there is some remaining emission from the blueshifted jet.
It is unclear if the lack of slow wind in the 2010 spectrum is real or an side-effect of the low spectral resolution of X-shooter.
During the 2022 burst (red line), there appears to be weak emission from slow winds, and, by the time of the post-burst spectrum (blue line), the emission from this slow wind had strengthened.
\citet{Banzatti2019_ApJ87076B} found that sources with higher \lacc{} have narrower low-velocity components.
Thus, it is possible that the low spectral resolution diluted a narrow component, which broadened as the \lacc{} decreased making it detectable.

\subsection{Accretion luminosity calculation}\label{ss:lacc}
Using the emission lines, we found that for each epoch, the values of \lacc{} per emission line are in agreement with each other within a factor of a few when considering the uncertainties.
In addition, we found that the pre- and post-outburst spectra resulted in the same values of \lacc{}, indicating that the outburst indeed had finished by late 2022 July.
The outburst spectrum resulted in \lacc{} values higher than the 2010 and 2022 post-outburst spectra by factors of 11$-$380 and 9$-$250 (depending on the line), respectively.
Based on the results of our slab modeling, the \lacc{} increased by factors of 5 and 9 when compared to the 2010 and 2022 quiescent spectra, respectively.
The resulting values of \lacc{} can be found in Table~\ref{table:macc} and we present a visual comparison of the \lacc{} values calculated with both methods in Fig.~\ref{fig:lacc}.
For the quiescent spectra, most of the emission lines indicated lower values of \lacc{} than the slab model, and only a few lines are in agreement with the slab model within an factor of 2$\sigma$.
For the outburst spectrum, more lines are in agreement with the slab model.

It is not clear what causes the discrepancy between the two methods.
We would naively expect that the coefficients found by \citet{Alcala2017_AA600A20A} would apply for \exl{}, even when in outburst, for multiple reasons.
First, the sample analyzed to find these relationships consisted of multiple T~Tauri stars including \exl{} itself.
Second, the stars in their sample were considered to be in quiescence but even if they were experiencing EXor-type outbursts, the stars would be still experiencing magnetospheric accretion albeit at a higher rate \citep{SiciliaAguilar2015_AA580A82S}.
Third, the coefficients are the best-fit parameters of a linear fit found between the logarithm of the line luminosity (for each emission line) and the logarithm of the accretion luminosity, which was determined using the same slab model fitting procedure we used.
Fourth, the line luminosities derived from our spectra are well within the range used by \citet{Alcala2017_AA600A20A} to determine these coefficients (see their Figures E.1 through E.6).
Therefore, we suspect that these differences can be attributed to the computation of the line fluxes because this is highly sensitive to how the continuum-subtracted spectrum was constructed and what spectral range was used for each line.
Indeed, \exl{} has a high \macc{}, even when in quiescence, thus finding its continuum level is not trivial for shorter wavelengths, and it is a line-rich star, which causes the accretion-tracing emission lines to be contaminated by other non-related lines.
In addition, as mentioned earlier, when the star was in outburst the line-blending is of higher concern because the lines became broader, and any lines that were previously too weak to be detected were now stronger and possibly contaminating the accretion-tracing lines.
These complications motivated us to only use the results from the slab modeling for the remainder of the discussion.

\subsection{Mass accretion rate}\label{ss:macc}
The transformation of \lacc{} to \macc{} depends on the assumptions of stellar radius and mass.
For this discussion we will focus on the \macc{} values calculated using the stellar parameters by \citet{Sipos2009_AA507881S}, and the accretion and stellar values used can be found in Table~\ref{table:macc}.
Afterwards we will briefly discuss how our conclusions would change if we had chosen slightly different values for the stellar properties.

Our estimations of \macc{} during quiescence are two orders of magnitude higher than was estimated previously by \citet{Sipos2009_AA507881S} and \citet{SiciliaAguilar2012_AA544A93S}.
Both studies used the relationship between 10\% width of H$\alpha$ and the \macc{} as determined by \citet{Natta2004_AA416179N}, and obtained values in the order of 10$^{-10}$\,M$_\odot$\,yr$^{-1}$.
We verified that the discrepancies between their estimates and ours are not due to observational differences (e.g.\ signal-to-noise ratio and spectral resolution) by using the \citet{Natta2004_AA416179N} relationship with our H$\alpha$ line profile.
The 10\% widths for our three H$\alpha$ measurements are 320\,km\,s$^{-1}$, 470\,km\,s$^{-1}$, and 360\,km\,s$^{-1}$ for the 2010 May, 2022 March and 2022 July, respectively, which would indicate \macc{} values of \sci{1.6}{-10}\,M$_\odot$\,yr$^{-1}$, \sci{4.6}{-9}\,M$_\odot$\,yr$^{-1}$, and \sci{3.8}{-10}\,M$_\odot$\,yr$^{-1}$, respectively.
Therefore, we confirm that using the 10\% width of H$\alpha$ on \exl{} would underestimate the \macc{}.
\citet{Sipos2009_AA507881S} also calculated \lacc{} and \macc{} based on the Pa$\beta$ line luminosity via the relationship by \citet{Muzerrolle1998_AJ1162965M}, and found a similar value as with the H$\alpha$ method.
The \lacc{} calculated by \citet{Sipos2009_AA507881S} using Pa$\beta$ is an order of magnitude lower than our lowest \lacc{} estimate from the emission lines and two orders lower than our estimate using the slab model (see Table~\ref{table:macc}).
These differences indicate that, when possible, one should aim to obtain more than one or two lines to determine the \macc{}, and in the case this is not possible then to consider the estimated \macc{} as highly uncertain.

Our value of \lacc{} is in agreement with that estimated by \citet{Alcala2017_AA600A20A}, which is expected as both estimates were done using the same spectra with the only difference being the slit-loss correcting factor, and the \macc{} values are in disagreement by a factor of a few but this can be explained by the differences in distance to \exl{} and the stellar properties assumed during this calculation.
These earlier estimates of \macc{}, including the 10\% widths of H$\alpha$, had suggested that \exl{} is a young star with one of the lowest mass accretion rates. However, our \macc{} results place \exl{} as one of the strongest accretors even when in quiescence \citep[e.g.][]{Alcala2017_AA600A20A,Banzatti2019_ApJ87076B}.

We can use our new estimate of the quiescent \macc{} to determine how strong the 2008 outburst was.
For that event, \citet{Juhasz2012_ApJ744118J} determined an outbursting \macc{} value of \sci{2}{-7}\,M$_\odot$\,yr$^{-1}$ using a Br$\gamma$ measurement taken $\sim$83\,days after the outburst reached its peak.
However, their spectrum was taken when \exl{} had dimmed by $\sim$2 magnitudes (see their Fig.\ 1) and they assumed an extinction $A_V$ of 0\,mag so we consider their value of \macc{} as a lower limit.
Indeed, their estimate of \macc{} for the 2008 outburst and is comparable to our estimates for the 2022 burst.
We obtained a new estimate using the methodology explained in Sect.\ \ref{ss:acc_curve} and anchoring our best-fit slab models to the photometry reported by \citet{Juhasz2012_ApJ744118J} in their Table\ 1.
This resulted in \macc{} $\approx$ \sci{(8.96 \pm 1.28)}{-7}\,M$_\odot$\,yr$^{-1}$ for the 2008 outburst.
Therefore, we find that this powerful event was due to an increase of the \macc{} by a factor of 52 and 29 when compared to our 2010 and 2022 quiescent spectra, respectively.
These ratios are comparable to what has been found for other EXor type stars \citep[e.g.][]{SiciliaAguilar2017_AA607A127S,Giannini2020_AA637A83G}.

As can be seen in Fig.\ \ref{fig:macc_curve}, during the peak of its 2022 burst, \exl{} reached \macc{} values 7 times higher than its quiescent level.
As we have shown, the 2022 burst was weaker and shorter than the 2008 outburst by a factor of a few, which is similar to the factor of 2 found by \citet{SiciliaAguilar2017_AA607A127S} between the outburst and burst of ASASSN-13db.

Our analysis has a couple of caveats to be considered when comparing our results to those from other works.
First, we assume that the extinction is $A_V=1.1$, when some studies have shown that this value could be as low as 0.0 \citep{Sipos2009_AA507881S} or 0.1 \citep{Wang2023_ApJsubmitted}, and so our \macc{} could be overestimated.
We tested this overestimation by running our slab-model fitting routine fixing $A_V=0$ and found that the \macc{} of all three epochs decreased by a factor of 4--5.
Therefore, our estimations of how \macc{} increased from quiescence are consistent but the impact of our choice of $A_V$ should be acknowledged when comparing our \macc{} values with those calculated for other stars.
Second, our choice of the stellar parameters and distance.
The choice of these parameters can affect the value of \macc{} by a factor of a few, as shown by our slab-model fitting results in Sect.~ \ref{ss:macc_lines_slab}.
Thus, when comparing our values with other analyses of \exl{}, these stellar parameters (radius and mass) and the distance to the star must be taken into account.

However, even though bursts are weaker than outbursts, they occur more frequently and so could also have a significant impact of the young star.

\subsection{Significance of one burst compared to other ones.}
Based on both its duration and amplitude, the 2022 burst is similar to the previous bursts seen in \exl{}.
Therefore, we can calculate the mass accreted by this event and apply this estimate to similar outbursts to obtain a rough estimate of the mass that has been accreted from the circumstellar disk during the outbursts.
For the discussion below, we assume that the disk of \exl{} has a total mass of 0.01\,M$_\odot$ \citep{Hales2018_ApJ859111H,White2020_ApJ90437W}.

We calculated the total mass accreted during the 2022 burst by integrating the \emph{accretion curve} found in Fig.~\ref{fig:macc_curve}.
Between 2022 February 14 and 2022 May 28, \exl{} accreted \sci{2.90}{-8}\,M$_\odot$ (0.8 Lunar masses).
To estimate how much more mass it was accreted due to the burst, we took the quiescent \macc{} value and multiplied it by the length of the burst (103 days) resulting in \sci{1.14}{-8}\,M$_\odot$ (0.3 Lunar masses).
Conversely, the excess mass accreted due to the burst was $M_b =$ \sci{1.76}{-8}\,M$_\odot$ (0.5 Lunar masses).
Therefore, during the burst, \exl{} accreted more than twice of the mass that would have during quiescence, thus, suggesting that a young star that experiences multiple bursts will deplete its circumstellar material faster than a non-outbursting one.

As mentioned above, finding the \macc{} history during the 2008 outburst is not an easy task, and thus, estimating the total mass accreted during this event is highly uncertain.
Our two estimates from Sect.\ \ref{ss:macc} are uncertain, so we will consider that the 2008 outburst reached a \macc{} value 40 times higher than the quiescent level and, thus, that it accreted $M_B =$ \sci{4.4}{-7}\,M$_\odot$ = 0.15 M$_\oplus$= 11.9 Moon masses in total.

Over the past 30 years, \exl{} has gone through 10 bursts and the powerful 2008 outburst. 
If we assume the 10 bursts accreted a similar amount of mass as the 2022 burst then the total mass accreted during three decades is $(M_q \times 30) + (M_b \times 10) + M_B =$ \sci{1.8}{-6}\,M$_\odot$.
This indicates that the outbursts were responsible for 34\% of the total mass accreted in the considered time span, of which 10\% and 24\% were the product of the bursts and the outburst, respectively.
If we consider that \exl{} lacks an envelope to replenish the disk material \citep{Sipos2009_AA507881S} and assume that the accretion in \exl{} will remain as active as it has been for these last three decades, it will deplete the total disk mass in 160\,000\,yr.

Nonetheless, we know from our full light curve (Fig.\ \ref{fig:full_lightcurve}) that, when compared with the 130 years, the accretion in \exl{} has been particularly active in these last three decades.
The pre-1990 data of the light curve is not as high quality as the more recent measurements, so estimating the total mass accreted during this time is still more uncertain.
Between 1950 and 1990, \citet{Bateson1991_PVSS1649B} reported several brightenings in the light curve, including the powerful 1955 outburst, and our inspection of the light curve indicates that during these decades there were 4 bursts.
The earlier data are harder to assess.
We find that there were at least 5 bursts in the light curve of \citet{McLaughlin1946_AJ52109M} plus the 1944 powerful outburst that had not been reported previously.
Our final tally of bursts and outbursts indicate that in the past 130 years there were at least 19 bursts and 3 outbursts, making the total accreted mass $(M_q \times 130) + (M_b \times 19) + (M_B \times 3) =$ \sci{6.9}{-6}\,M$_\odot$.
A similar behavior in the \macc{} would consume the disk mass in 190\,000\,yr.
Finally, if we assume that \exl{} will not experience another burst or outburst then at its quiescent \macc{}, it will deplete its mass reservoir in 250\,000\,yr.

As mentioned earlier, this estimate is uncertain due to the multiple assumptions such as the constant quiescent \macc{} (which might not be stable if we consider that the ``quiescent'' brightness of the full light curve has changed by $\sim$1 magnitude between e.g.\ 1980 and 2020) and the lower limit of detected bursts and outbursts (due to the long gaps in the photometric coverage).
Still, all estimates suggest that \exl{} is close to spending its disk mass and will soon move to its next evolutionary phase.

\section{Summary \& Conclusions}\label{sec:theend}
\exl{} started experiencing a new brightening event in February 2022 and by late May it had returned to its normal brightness levels.
We carried out a multi-filter photometric monitoring and obtained two spectroscopic measurements, and found that this brightening was the result of an increase in the mass accretion rate.
Our main findings are:

\begin{enumerate}
  \item Using $g$ band photometry we found that the burst started in the middle of February 2022 and ended in late May. \exl{} brightened by $\sim$2\,mag at its peak, and experienced brightening and dimming rates of 0.036\,mag\,day$^{-1}$ and 0.026\,mag\,day$^{-1}$, respectively.
  \item We saw some periodicity in the light curve and analyzed it using Lomb-Scargle diagrams at different stages of the burst. However, we did not find significant changes in the period of these brightness fluctuations. We find a $\sim$7.4-day period in \exl{} during and post-outburst, as well as during quiescence, consistent with previous studies.
  \item Using color-color and color-magnitude diagrams, we found that the brightening was not caused by changes in the extinction.
  \item The spectra of \exl{} are crowded with lines due to neutral and ionized atoms, and these became broader and stronger during the burst. We briefly discussed the evolution of [O~I] line which traces jets and winds, and found differences in the line profile that are tentatively due to the burst but can also be explained by the spectral resolution of our observations.
  \item Using different emission lines to calculate the \lacc{}, we derived a wide range of values for each epoch, and these differ from the values calculated with the slab model. Using the emission lines opens the window to higher uncertainties, in particular with the calculation of the continuum, and thus, we recommend using the slab model when possible or more than one emission line to calculate the \lacc{}.
  \item Based on the slab model fitting, during the burst, the \macc{} increased by a factor of $\sim$7. The total mass accreted during the burst was 0.8 Lunar masses, which is $\sim$1.5 times more than would have been accreted in the same length of time with the quiescent level of \macc{}.
  \item We have estimated that, for the past 130 years, 30\% of the accreted mass has been the product of bursts and outbursts. Therefore, we find that these events can significantly accelerate the evolution of a young stellar object.
  \item The combination of the mass of the disk surrounding \exl{} and an extrapolation of the accretion history based on 130 years of observations, suggests that, in the absence of significant replenishment, the disk will be depleted in a few hundred thousand years, and thus, it is close to reaching its next evolutionary stage. 
\end{enumerate}

The 2022 burst of \exl{} provided great insight into the short-term changes brought by a single accretion burst, and shed some light into the long-term effects that multiple bursts and outbursts can have on a young star.
However, several questions remain regarding these accretion bursts and outbursts \citep{Fischer2023_PPVII}.
In-depth multi-wavelength observational studies of future bursts and outbursts are needed to begin answering these questions.

\begin{acknowledgements}
This project has received funding from the European Research Council (ERC) under the European Union’s Horizon 2020 research and innovation programme under grant agreement No 716155 (SACCRED).
This work has been funded by the European Union under the European Union’s Horizon Europe Research \& Innovation Programme 101039452 (WANDA).
INAF co-authors acknowledge the support of PRIN-INAF 2019 "Spectroscopically Tracing the Disk Dispersal Evolution (STRADE)" and by the Large Grant INAF 2022 YODA (YSOs Outflows, Disks and Accretion: towards a global framework for the evolution of planet forming systems)
Zs.M.Sz.\ acknowledges funding from a St Leonards scholarship from the University of St Andrews. 
Zs.M.Sz.\ is a member of the International Max Planck Research School (IMPRS) for Astronomy and Astrophysics at the Universities of Bonn and Cologne.
Zs.N.\ was supported by the János Bolyai Research Scholarship of the Hungarian Academy of Sciences.
We acknowledge support from the ESA PRODEX contract nr.\ 4000132054.
We thank the REM staff for the observational support.
This work makes use of observations from the Las Cumbres Observatory global telescope network.
Based on observations collected at the European Southern Observatory under ESO programmes 085.C-0764(A), 108.23N8.001, and 109.24F7.001.
\end{acknowledgements}

\bibliographystyle{aa} 
\bibliography{references}

\begin{appendix}
\section{The 1944 outburst of EX~Lupi}\label{app:lc}
We searched for \exl{} photometry on the  Digital Access to a Sky Century at Harvard (DASCH) project website\footnote{\url{https://library.cfa.harvard.edu/dasch}}.
Their light curve tool offers four different options for the input catalog to be used as reference for the magnitude estimation.
We selected the APASS Input Catalog, which offers \textit{Visible} magnitudes and does not need a correction factor with respect to the values obtained from AAVSO.
The full light curve can be seen in Fig.\ \ref{fig:full_lightcurve}.
It roughly covers from 1893 to 1953, then 1970 to 1973, and from 1978 to 1989, with the earliest plate taken in 1893 June 24.

\begin{figure}[!hb]
  \begin{minipage}[c]{\textwidth}
  \centering
  \includegraphics[width=18cm]{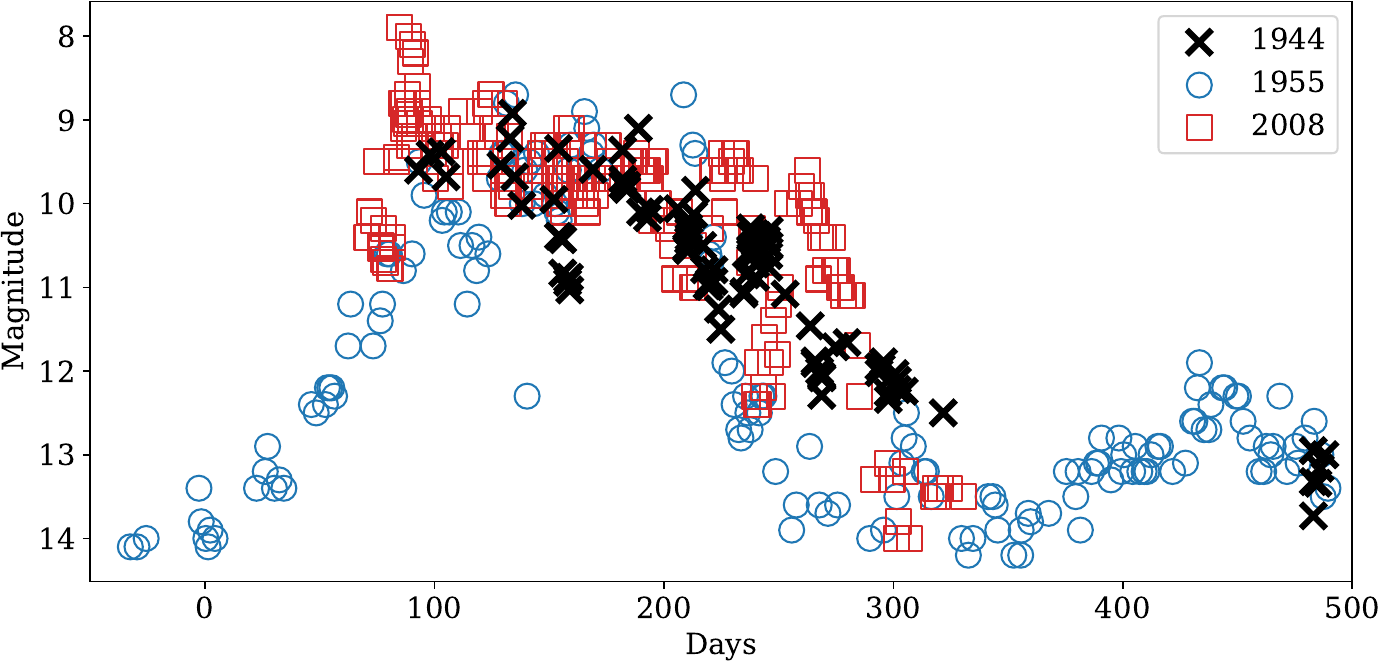}
  \caption{The three known powerful outbursts of \exl{}. They have been shifted to approximately align the length of their outbursts.\label{fig:three_amigos}}
  \end{minipage}
\end{figure}

In Fig.\ \ref{fig:three_amigos} we show the three known powerful outbursts of \exl{}.
The sampling of the light curve for the three events is different.
We lack the data points to know when the 1944 outburst began and when it went back to quiescence, the 1955 outburst is the the best sampled event with an adequate coverage of the brightening and dimming phases, and the 2008 outburst has a few data points still in brightening but we lack information of when it started to brighten.
Nevertheless, we have shifted the three events to approximately align them by assuming that the first data point of the 1944 event is just after the outburst reached its peak.
The MJD Day 0 are 31400, 35250 and 54410 for the 1944, 1955 and 2008 outbursts, respectively.

The three outbursts show significant variability in their bright stages, dimming by as much as 2$\sim$3\,mag.
The 1955 outburst has been the shortest, lasting for $\sim$250 days, while it is likely that the 1944 outburst was the longest by lasting $\gtrsim$230 days.
The three outbursts reached comparable magnitudes in their heightened stages, indicative of similar physical conditions for each outburst.
However, there is spectroscopy only for the 2008 event and it is of moderate quality.
Therefore, the determination of these physical properties shall be pursued in a future powerful outburst.

\section{Line fluxes and line luminosities}
Here we present the line fluxes and line luminosities for each accretion-tracing emission line per epoch.

\begin{table*}
  \caption{Line flux and line luminosities for the 2010 May 04 observations. Their calculation in explained in Sect.~\ref{ss:acclum}.\label{table:lflx_llum_2010}}
\centering
\begin{tabular}{lrccccccccc}
\hline\hline
Line & \multicolumn{1}{c}{$\lambda$} & $L_\mathrm{flx}$ & $L_\mathrm{lum}$\\
     & \multicolumn{1}{c}{[nm]} & [erg s$^{-1}$ cm$^{-2}$] & [L$_\odot$] \\
\hline
H15         & 371.20  & \sci{3.70}{-14} $\pm$ \sci{7.76}{-16} & \sci{2.78}{-5} $\pm$ \sci{5.82}{-7}\\
H14         & 372.19  & \sci{4.78}{-14} $\pm$ \sci{5.82}{-16} & \sci{3.59}{-5} $\pm$ \sci{4.37}{-7}\\
H13         & 373.44  & \sci{7.59}{-14} $\pm$ \sci{5.62}{-16} & \sci{5.70}{-5} $\pm$ \sci{4.22}{-7}\\
H12         & 375.02  & \sci{1.12}{-13} $\pm$ \sci{7.76}{-16} & \sci{8.41}{-5} $\pm$ \sci{5.83}{-7}\\
H11         & 377.06  & \sci{1.81}{-13} $\pm$ \sci{1.02}{-15} & \sci{1.36}{-4} $\pm$ \sci{7.68}{-7}\\
H10         & 379.79  & \sci{2.20}{-13} $\pm$ \sci{9.55}{-16} & \sci{1.65}{-4} $\pm$ \sci{7.17}{-7}\\
H9          & 383.54  & \sci{3.24}{-13} $\pm$ \sci{1.08}{-15} & \sci{2.43}{-4} $\pm$ \sci{8.08}{-7}\\
H8          & 388.90  & \sci{5.18}{-13} $\pm$ \sci{1.19}{-15} & \sci{3.89}{-4} $\pm$ \sci{8.91}{-7}\\
Ca II K     & 393.37  & \sci{4.90}{-13} $\pm$ \sci{9.01}{-16} & \sci{3.68}{-4} $\pm$ \sci{6.77}{-7}\\
Ca II H     & 396.85  & \sci{4.24}{-13} $\pm$ \sci{6.58}{-16} & \sci{3.18}{-4} $\pm$ \sci{4.94}{-7}\\
H$\epsilon$ & 397.01  & \sci{4.27}{-13} $\pm$ \sci{9.15}{-16} & \sci{3.21}{-4} $\pm$ \sci{6.87}{-7}\\
He I        & 402.62  & \sci{4.27}{-14} $\pm$ \sci{4.99}{-16} & \sci{3.21}{-5} $\pm$ \sci{3.75}{-7}\\
H$\delta$   & 410.17  & \sci{6.25}{-13} $\pm$ \sci{1.03}{-15} & \sci{4.69}{-4} $\pm$ \sci{7.74}{-7}\\
H$\gamma$   & 434.05  & \sci{6.88}{-13} $\pm$ \sci{9.10}{-16} & \sci{5.17}{-4} $\pm$ \sci{6.83}{-7}\\
He I        & 447.15  & \sci{7.31}{-14} $\pm$ \sci{3.77}{-16} & \sci{5.49}{-5} $\pm$ \sci{2.83}{-7}\\
He II       & 468.58  & \sci{3.29}{-14} $\pm$ \sci{2.75}{-16} & \sci{2.47}{-5} $\pm$ \sci{2.06}{-7}\\
He I        & 471.31  & \sci{1.08}{-14} $\pm$ \sci{2.01}{-16} & \sci{8.09}{-6} $\pm$ \sci{1.51}{-7}\\
H$\beta$    & 486.13  & \sci{7.81}{-13} $\pm$ \sci{8.24}{-16} & \sci{5.86}{-4} $\pm$ \sci{6.19}{-7}\\
He I Fe I   & 492.19  & \sci{6.61}{-14} $\pm$ \sci{3.08}{-16} & \sci{4.96}{-5} $\pm$ \sci{2.31}{-7}\\
He I        & 501.57  & \sci{1.77}{-14} $\pm$ \sci{1.68}{-16} & \sci{1.33}{-5} $\pm$ \sci{1.26}{-7}\\
He I        & 587.56  & \sci{9.45}{-14} $\pm$ \sci{3.14}{-16} & \sci{7.10}{-5} $\pm$ \sci{2.36}{-7}\\
Na I        & 589.00  & \sci{1.99}{-14} $\pm$ \sci{1.65}{-16} & \sci{1.49}{-5} $\pm$ \sci{1.24}{-7}\\
Na I        & 589.59  & \sci{1.06}{-14} $\pm$ \sci{1.22}{-16} & \sci{7.96}{-6} $\pm$ \sci{9.17}{-8}\\
H$\alpha$   & 656.28  & \sci{1.69}{-12} $\pm$ \sci{5.31}{-16} & \sci{1.27}{-3} $\pm$ \sci{3.99}{-7}\\
He I        & 667.82  & \sci{3.45}{-14} $\pm$ \sci{1.80}{-16} & \sci{2.59}{-5} $\pm$ \sci{1.35}{-7}\\
He I        & 706.52  & \sci{2.24}{-14} $\pm$ \sci{1.21}{-16} & \sci{1.68}{-5} $\pm$ \sci{9.10}{-8}\\
O I         & 777.31  & \sci{2.73}{-14} $\pm$ \sci{1.47}{-16} & \sci{2.05}{-5} $\pm$ \sci{1.10}{-7}\\
O I         & 844.64  & \sci{1.54}{-14} $\pm$ \sci{1.51}{-16} & \sci{1.15}{-5} $\pm$ \sci{1.13}{-7}\\
Ca II       & 849.80  & \sci{1.64}{-13} $\pm$ \sci{1.61}{-16} & \sci{1.23}{-4} $\pm$ \sci{1.21}{-7}\\
Ca II       & 854.21  & \sci{1.80}{-13} $\pm$ \sci{1.56}{-16} & \sci{1.35}{-4} $\pm$ \sci{1.17}{-7}\\
Ca II       & 866.21  & \sci{1.60}{-13} $\pm$ \sci{1.71}{-16} & \sci{1.20}{-4} $\pm$ \sci{1.28}{-7}\\
Pa10        & 901.49  & \sci{1.44}{-14} $\pm$ \sci{1.05}{-16} & \sci{1.08}{-5} $\pm$ \sci{7.88}{-8}\\
Pa9         & 922.90  & \sci{3.29}{-14} $\pm$ \sci{1.51}{-16} & \sci{2.47}{-5} $\pm$ \sci{1.14}{-7}\\
Pa8         & 954.60  & \sci{7.26}{-14} $\pm$ \sci{2.00}{-16} & \sci{5.45}{-5} $\pm$ \sci{1.50}{-7}\\
Pa$\delta$  & 1004.94 & \sci{7.69}{-14} $\pm$ \sci{3.81}{-16} & \sci{5.77}{-5} $\pm$ \sci{2.86}{-7}\\
Pa$\gamma$  & 1093.81 & \sci{1.07}{-13} $\pm$ \sci{2.36}{-16} & \sci{8.07}{-5} $\pm$ \sci{1.77}{-7}\\
Pa$\beta$   & 1281.81 & \sci{8.77}{-14} $\pm$ \sci{1.98}{-16} & \sci{6.58}{-5} $\pm$ \sci{1.49}{-7}\\
Br$\gamma$  & 2166.12 & \sci{1.52}{-14} $\pm$ \sci{6.74}{-17} & \sci{1.14}{-5} $\pm$ \sci{5.06}{-8}\\
\hline
\end{tabular}
\end{table*}

\begin{table*}
  \caption{Line flux and line luminosities for the 2022 March 27 observations. Their calculation in explained in Sect.~\ref{ss:acclum}.\label{table:lflx_llum_202203}}
\centering
\begin{tabular}{lrccccccccc}
\hline\hline
Line & \multicolumn{1}{c}{$\lambda$} & $L_\mathrm{flx}$ & $L_\mathrm{lum}$\\
     & \multicolumn{1}{c}{[nm]} & [erg s$^{-1}$ cm$^{-2}$] & [L$_\odot$] \\
\hline
H15         & 371.20  & \sci{7.42}{-13} $\pm$ \sci{1.24}{-15} & \sci{5.58}{-4} $\pm$ \sci{9.34}{-7}\\
H14         & 372.19  & \sci{2.00}{-12} $\pm$ \sci{1.01}{-15} & \sci{1.51}{-3} $\pm$ \sci{7.60}{-7}\\
H13         & 373.44  & \sci{2.74}{-12} $\pm$ \sci{1.08}{-15} & \sci{2.06}{-3} $\pm$ \sci{8.10}{-7}\\
H12         & 375.02  & \sci{2.49}{-12} $\pm$ \sci{1.23}{-15} & \sci{1.87}{-3} $\pm$ \sci{9.23}{-7}\\
H11         & 377.06  & \sci{4.90}{-12} $\pm$ \sci{1.59}{-15} & \sci{3.68}{-3} $\pm$ \sci{1.19}{-6}\\
H10         & 379.79  & \sci{7.38}{-12} $\pm$ \sci{1.66}{-15} & \sci{5.54}{-3} $\pm$ \sci{1.25}{-6}\\
H9          & 383.54  & \sci{3.22}{-12} $\pm$ \sci{1.24}{-15} & \sci{2.42}{-3} $\pm$ \sci{9.29}{-7}\\
H8          & 388.90  & \sci{1.48}{-11} $\pm$ \sci{2.03}{-15} & \sci{1.11}{-2} $\pm$ \sci{1.52}{-6}\\
Ca II K     & 393.37  & \sci{7.80}{-11} $\pm$ \sci{2.63}{-15} & \sci{5.86}{-2} $\pm$ \sci{1.98}{-6}\\
Ca II H     & 396.85  & \sci{3.21}{-11} $\pm$ \sci{1.55}{-15} & \sci{2.41}{-2} $\pm$ \sci{1.16}{-6}\\
H$\epsilon$ & 397.01  & \sci{1.50}{-11} $\pm$ \sci{1.32}{-15} & \sci{1.13}{-2} $\pm$ \sci{9.92}{-7}\\
He I        & 402.62  & \sci{1.55}{-12} $\pm$ \sci{8.28}{-16} & \sci{1.16}{-3} $\pm$ \sci{6.22}{-7}\\
H$\delta$   & 410.17  & \sci{1.96}{-11} $\pm$ \sci{1.73}{-15} & \sci{1.47}{-2} $\pm$ \sci{1.30}{-6}\\
H$\gamma$   & 434.05  & \sci{2.32}{-11} $\pm$ \sci{1.57}{-15} & \sci{1.74}{-2} $\pm$ \sci{1.18}{-6}\\
He I        & 447.15  & \sci{1.61}{-12} $\pm$ \sci{5.34}{-16} & \sci{1.21}{-3} $\pm$ \sci{4.01}{-7}\\
He II       & 468.58  & \sci{1.26}{-12} $\pm$ \sci{5.29}{-16} & \sci{9.46}{-4} $\pm$ \sci{3.97}{-7}\\
He I        & 471.31  & \sci{2.73}{-13} $\pm$ \sci{3.14}{-16} & \sci{2.05}{-4} $\pm$ \sci{2.36}{-7}\\
H$\beta$    & 486.13  & \sci{3.70}{-11} $\pm$ \sci{1.40}{-15} & \sci{2.78}{-2} $\pm$ \sci{1.05}{-6}\\
He I Fe I   & 492.19  & \sci{6.36}{-12} $\pm$ \sci{6.93}{-16} & \sci{4.78}{-3} $\pm$ \sci{5.20}{-7}\\
He I        & 501.57  & \sci{1.11}{-12} $\pm$ \sci{3.21}{-16} & \sci{8.33}{-4} $\pm$ \sci{2.41}{-7}\\
He I        & 587.56  & \sci{2.83}{-12} $\pm$ \sci{4.88}{-16} & \sci{2.13}{-3} $\pm$ \sci{3.66}{-7}\\
Na I        & 589.00  & \sci{3.68}{-12} $\pm$ \sci{5.09}{-16} & \sci{2.77}{-3} $\pm$ \sci{3.82}{-7}\\
Na I        & 589.59  & \sci{1.35}{-12} $\pm$ \sci{3.70}{-16} & \sci{1.01}{-3} $\pm$ \sci{2.78}{-7}\\
H$\alpha$   & 656.28  & \sci{1.09}{-1} $\pm$ \sci{1.03}{-15}  & \sci{8.20}{-2} $\pm$ \sci{7.71}{-7}\\
He I        & 667.82  & \sci{1.28}{-12} $\pm$ \sci{3.80}{-16} & \sci{9.65}{-4} $\pm$ \sci{2.85}{-7}\\
He I        & 706.52  & \sci{7.22}{-13} $\pm$ \sci{1.83}{-16} & \sci{5.42}{-4} $\pm$ \sci{1.37}{-7}\\
O I         & 777.31  & \sci{1.23}{-12} $\pm$ \sci{1.78}{-16} & \sci{9.26}{-4} $\pm$ \sci{1.33}{-7}\\
O I         & 844.64  & \sci{3.79}{-12} $\pm$ \sci{2.03}{-16} & \sci{2.85}{-3} $\pm$ \sci{1.53}{-7}\\
Ca II       & 849.80  & \sci{4.33}{-11} $\pm$ \sci{4.27}{-16} & \sci{3.25}{-2} $\pm$ \sci{3.21}{-7}\\
Ca II       & 854.21  & \sci{4.62}{-11} $\pm$ \sci{4.31}{-16} & \sci{3.47}{-2} $\pm$ \sci{3.23}{-7}\\
Ca II       & 866.21  & \sci{3.97}{-11} $\pm$ \sci{4.09}{-16} & \sci{2.98}{-2} $\pm$ \sci{3.07}{-7}\\
Pa10        & 901.49  & \sci{8.21}{-13} $\pm$ \sci{1.09}{-16} & \sci{6.16}{-4} $\pm$ \sci{8.17}{-8}\\
Pa9         & 922.90  & \sci{1.56}{-12} $\pm$ \sci{2.12}{-16} & \sci{1.17}{-3} $\pm$ \sci{1.59}{-7}\\
Pa8         & 954.60  & \sci{2.70}{-12} $\pm$ \sci{2.06}{-16} & \sci{2.02}{-3} $\pm$ \sci{1.55}{-7}\\
Pa$\delta$  & 1004.94 & \sci{3.27}{-12} $\pm$ \sci{3.82}{-16} & \sci{2.46}{-3} $\pm$ \sci{2.87}{-7}\\
Pa$\gamma$  & 1093.81 & \sci{4.57}{-12} $\pm$ \sci{2.18}{-16} & \sci{3.43}{-3} $\pm$ \sci{1.64}{-7}\\
Pa$\beta$   & 1281.81 & \sci{6.79}{-12} $\pm$ \sci{2.19}{-16} & \sci{5.10}{-3} $\pm$ \sci{1.65}{-7}\\
Br$\gamma$  & 2166.12 & \sci{7.46}{-13} $\pm$ \sci{5.05}{-17} & \sci{5.60}{-4} $\pm$ \sci{3.79}{-8}\\
\hline
\end{tabular}
\end{table*}

\begin{table*}
  \caption{Line flux and line luminosities for the 2022 July 29 observations. Their calculation in explained in Sect.~\ref{ss:acclum}.\label{table:lflx_llum_202207}}
\centering
\begin{tabular}{lrccccccccc}
\hline\hline
Line & \multicolumn{1}{c}{$\lambda$} & $L_\mathrm{flx}$ & $L_\mathrm{lum}$\\
     & \multicolumn{1}{c}{[nm]} & [erg s$^{-1}$ cm$^{-2}$] & [L$_\odot$] \\
\hline
H15         & 371.20  & \sci{3.12}{-14} $\pm$ \sci{7.10}{-16} & \sci{2.35}{-5} $\pm$ \sci{5.33}{-7}\\
H14         & 372.19  & \sci{4.58}{-14} $\pm$ \sci{5.49}{-16} & \sci{3.44}{-5} $\pm$ \sci{4.12}{-7}\\
H13         & 373.44  & \sci{7.42}{-14} $\pm$ \sci{5.37}{-16} & \sci{5.57}{-5} $\pm$ \sci{4.03}{-7}\\
H12         & 375.02  & \sci{8.54}{-14} $\pm$ \sci{6.86}{-16} & \sci{6.41}{-5} $\pm$ \sci{5.15}{-7}\\
H11         & 377.06  & \sci{1.95}{-13} $\pm$ \sci{9.40}{-16} & \sci{1.47}{-4} $\pm$ \sci{7.06}{-7}\\
H10         & 379.79  & \sci{2.87}{-13} $\pm$ \sci{9.88}{-16} & \sci{2.15}{-4} $\pm$ \sci{7.42}{-7}\\
H9          & 383.54  & \sci{4.02}{-13} $\pm$ \sci{1.07}{-15} & \sci{3.02}{-4} $\pm$ \sci{8.02}{-7}\\
H8          & 388.90  & \sci{7.12}{-13} $\pm$ \sci{1.13}{-15} & \sci{5.35}{-4} $\pm$ \sci{8.50}{-7}\\
Ca II K     & 393.37  & \sci{7.46}{-13} $\pm$ \sci{8.80}{-16} & \sci{5.60}{-4} $\pm$ \sci{6.61}{-7}\\
Ca II H     & 396.85  & \sci{7.07}{-13} $\pm$ \sci{7.95}{-16} & \sci{5.31}{-4} $\pm$ \sci{5.97}{-7}\\
H$\epsilon$ & 397.01  & \sci{4.97}{-13} $\pm$ \sci{7.93}{-16} & \sci{3.73}{-4} $\pm$ \sci{5.95}{-7}\\
He I        & 402.62  & \sci{4.42}{-14} $\pm$ \sci{4.39}{-16} & \sci{3.32}{-5} $\pm$ \sci{3.29}{-7}\\
H$\delta$   & 410.17  & \sci{9.47}{-13} $\pm$ \sci{9.99}{-16} & \sci{7.11}{-4} $\pm$ \sci{7.50}{-7}\\
H$\gamma$   & 434.05  & \sci{1.04}{-12} $\pm$ \sci{8.84}{-16} & \sci{7.81}{-4} $\pm$ \sci{6.64}{-7}\\
He I        & 447.15  & \sci{8.84}{-14} $\pm$ \sci{3.88}{-16} & \sci{6.64}{-5} $\pm$ \sci{2.92}{-7}\\
He II       & 468.58  & \sci{3.83}{-14} $\pm$ \sci{2.74}{-16} & \sci{2.88}{-5} $\pm$ \sci{2.06}{-7}\\
He I        & 471.31  & \sci{1.27}{-14} $\pm$ \sci{2.19}{-16} & \sci{9.54}{-6} $\pm$ \sci{1.65}{-7}\\
H$\beta$    & 486.13  & \sci{1.10}{-12} $\pm$ \sci{8.45}{-16} & \sci{8.28}{-4} $\pm$ \sci{6.35}{-7}\\
He I Fe I   & 492.19  & \sci{7.00}{-14} $\pm$ \sci{2.95}{-16} & \sci{5.25}{-5} $\pm$ \sci{2.22}{-7}\\
He I        & 501.57  & \sci{2.12}{-14} $\pm$ \sci{1.95}{-16} & \sci{1.59}{-5} $\pm$ \sci{1.46}{-7}\\
He I        & 587.56  & \sci{1.16}{-13} $\pm$ \sci{3.15}{-16} & \sci{8.68}{-5} $\pm$ \sci{2.36}{-7}\\
Na I        & 589.00  & \sci{3.52}{-14} $\pm$ \sci{1.96}{-16} & \sci{2.64}{-5} $\pm$ \sci{1.47}{-7}\\
Na I        & 589.59  & \sci{2.20}{-14} $\pm$ \sci{1.86}{-16} & \sci{1.66}{-5} $\pm$ \sci{1.39}{-7}\\
H$\alpha$   & 656.28  & \sci{1.79}{-12} $\pm$ \sci{5.37}{-16} & \sci{1.35}{-3} $\pm$ \sci{4.03}{-7}\\
He I        & 667.82  & \sci{4.65}{-14} $\pm$ \sci{1.93}{-16} & \sci{3.49}{-5} $\pm$ \sci{1.45}{-7}\\
He I        & 706.52  & \sci{2.47}{-14} $\pm$ \sci{1.33}{-16} & \sci{1.86}{-5} $\pm$ \sci{1.00}{-7}\\
O I         & 777.31  & \sci{3.26}{-14} $\pm$ \sci{1.44}{-16} & \sci{2.45}{-5} $\pm$ \sci{1.08}{-7}\\
O I         & 844.64  & \sci{2.30}{-14} $\pm$ \sci{1.53}{-16} & \sci{1.73}{-5} $\pm$ \sci{1.15}{-7}\\
Ca II       & 849.80  & \sci{1.82}{-13} $\pm$ \sci{1.80}{-16} & \sci{1.37}{-4} $\pm$ \sci{1.35}{-7}\\
Ca II       & 854.21  & \sci{2.22}{-13} $\pm$ \sci{1.78}{-16} & \sci{1.67}{-4} $\pm$ \sci{1.33}{-7}\\
Ca II       & 866.21  & \sci{1.86}{-13} $\pm$ \sci{1.79}{-16} & \sci{1.40}{-4} $\pm$ \sci{1.34}{-7}\\
Pa10        & 901.49  & \sci{1.17}{-14} $\pm$ \sci{9.80}{-17} & \sci{8.77}{-6} $\pm$ \sci{7.36}{-8}\\
Pa9         & 922.90  & \sci{2.40}{-14} $\pm$ \sci{1.57}{-16} & \sci{1.80}{-5} $\pm$ \sci{1.18}{-7}\\
Pa8         & 954.60  & \sci{6.64}{-14} $\pm$ \sci{2.11}{-16} & \sci{4.99}{-5} $\pm$ \sci{1.58}{-7}\\
Pa$\delta$  & 1004.94 & \sci{7.23}{-14} $\pm$ \sci{4.05}{-16} & \sci{5.43}{-5} $\pm$ \sci{3.04}{-7}\\
Pa$\gamma$  & 1093.81 & \sci{9.86}{-14} $\pm$ \sci{2.52}{-16} & \sci{7.41}{-5} $\pm$ \sci{1.89}{-7}\\
Pa$\beta$   & 1281.81 & \sci{9.25}{-14} $\pm$ \sci{2.14}{-16} & \sci{6.95}{-5} $\pm$ \sci{1.61}{-7}\\
Br$\gamma$  & 2166.12 & \sci{1.14}{-14} $\pm$ \sci{6.74}{-17} & \sci{8.56}{-6} $\pm$ \sci{5.06}{-8}\\
\hline
\end{tabular}
\end{table*}

\end{appendix}

\end{document}